\documentclass[%
 reprint,
 amsmath,amssymb,
pra,
]{revtex4-2}

\usepackage{graphicx}
\usepackage{dcolumn}
\usepackage{bm}
\usepackage[english]{babel}
\usepackage{amsmath, amsthm, amssymb, amsfonts}
\usepackage{hyperref}
\usepackage{bbold}
\usepackage{physics}
\usepackage{xcolor}

\begin{document}

\title{Floquet engineering of tight-binding Hamiltonians in momentum space lattices}
\author{D. Ronco, F. Arrouas, N. Ombredane, E. Flament, Q. Levoy, B. Peaudecerf and D. Guéry-Odelin}

\affiliation{
 Laboratoire Collisions, Agrégats, Réactivité, FeRMI, Université de Toulouse,
CNRS, 118 Route de Narbonne, F-31062 Toulouse Cedex 09, France 
}%

\date{\today}

\begin{abstract}
Quantum simulation with ultracold atoms provides a versatile platform to emulate condensed-matter models. In particular, momentum-space lattices enable the realization of programmable tight-binding Hamiltonians. 
Here, we generalize this approach by exploiting quantum resonances of a periodically driven (shaken) rotor within the Floquet framework. Using first-order time-dependent perturbation theory, we derive analytical relations between the lattice modulation and the effective tight-binding parameters, and identify explicit solutions for several resonances. We further apply optimal-control techniques to enhance the multi-period Floquet fidelity and extend the accessible parameter regimes.
Experimentally, we implement this scheme with a Bose–Einstein condensate of rubidium-87 atoms in a dynamically modulated optical lattice. We demonstrate the simulation of the Rice–Mele model, including band-structure measurements and topological edge states, as well as momentum Bloch oscillations, and superlattice configurations with controlled periodicity.
Our results establish quantum resonances as a powerful resource for Floquet engineering of tight-binding models in momentum space.
\end{abstract}

\maketitle

\section{\label{sec:level1} Introduction }

To tackle the complexity of computing properties and dynamics of many-body quantum systems, quantum simulation resorts to highly tunable experimental quantum platforms that can emulate them~\cite{feynman_simulating_1982, lloyd_universal_1996, bloch_many-body_2008, georgescu_quantum_2014, gross_quantum_2017}. Over the last two decades, cold-atom experiments have developed a versatile toolkit to shape and control the environment of quantum gases \cite{lewenstein_ultracold_2007, Bloch_Nature_2012, schafer_tools_2020}. As a result, cold atoms can simulate paradigmatic condensed-matter models: band structures with Dirac points~\cite{soltan-panahi_multi-component_2011, Tarruell_Nature_2012, uehlinger_artificial_2013}, disordered systems displaying Anderson localization and the associated transition~\cite{Billy_Nature_2008, roati_anderson_2008, chabe_experimental_2008, Lemarie_PRL_2010,sanchez-palencia_disordered_2010,  Kondov_Science_2011, jendrzejewski_three-dimensional_2012,shapiro_cold_2012,  semeghini_measurement_2015, white_observation_2020}, as well as topological lattice models (e.g.\ Haldane~\cite{haldane_model_1988, Jotzu_Nature_2014}, Rice--Mele \cite{Rice_Mele_1982, thouless_quantization_1983, nakajima_topological_2016, lohse_thouless_2016, citro_thouless_2023}, and Hofstadter~\cite{jaksch_creation_2003, Aidelsburger_PRL_2013, miyake_realizing_2013}, along with kagome geometries \cite{jo_ultracold_2012, leung_interaction-enhanced_2020}). Beyond engineering these Hamiltonians, their topological properties have been probed directly, including measurements of the Zak phase~\cite{Zak_1989, Atala_NaturePhys_2013, cooper_topological_2019}, Chern numbers~\cite{Aidelsburger_NaturePhys_2015, asteria_measuring_2019}, and the observation of edge states~\cite{Mancini_Science_2015,Stuhl_Science_2015, celi_synthetic_2014, sompet2022realizing, chalopin_2020probing}. In particular, the ability to modulate parameters in time enables Floquet engineering, e.g. the realization of effective tunnelling amplitudes in lattices~\cite{lignier_dynamical_2007, Goldman_PRX_2014, bukov_universal_2015, weitenberg_tailoring_2021, rudner_band_2020, eckardt_atomic_2017} or of artificial gauge fields~\cite{lin_synthetic_2009, Aidelsburger_PRL_2011,dalibard_artificial_2011, Kennedy_Nature_2015,goldman_light-induced_2014}.

Although most condensed-matter models are formulated in real (position) space, cold-atom platforms also enable faithful implementations of lattices in momentum space. Quantum simulations using momentum-space lattices (MSLs) and the Bragg resonance condition~\cite{Kozuma_PRL_1999} have been implemented to emulate the dynamics of a one-dimensional tight-binding model with uniform on-site energies and arbitrarily programmable (next-)nearest-neighbor tunnelling coefficients~\cite{Gadway_PRA_2015,Meier_PRA_2016}. This toolbox has enabled studies of topological models such as the Su-Schrieffer-Heeger model~\cite{Xie_Nature_2019} and synthetic magnetic ladders~\cite{An_ScienceAdv_2017}, of disordered systems~\cite{Meier_PRA_2019}, and of interaction effects in momentum-space lattices~\cite{An_PRL_2018}.
Momentum space has also been exploited to emulate disorder, in particular with the rich physics of the kicked Rotor dynamic \cite{Chirikov_SovSci_1981}. A standard example is provided by the study of Anderson localization in momentum space for which a momentum-space lattice with quasi-random on-site energies is realized \cite{Chirikov_SovSci_1981,Moore_PRL_1995}. 

Interestingly, the kicked rotor also exhibits quantum resonances~\cite{IzrailShepe_THEORMath_1980}, which occur for specific values of the Floquet period. Such resonances have been observed experimentally with Bose–Einstein condensates~\cite{Ryu_PRL_2006}. However, these resonant regimes have not yet been exploited directly as quantum simulators operating in momentum space. In this work, we theoretically and experimentally investigate a shaken rotor (SR) model at a quantum resonance as a platform for quantum simulation in momentum space. The SR model generalizes the kicked rotor, offering a rich variety of control possibilities through its additional degrees of freedom. We also show that the Bragg scattering regime naturally emerges as a specific quantum resonance of the shaken rotor. By combining quantum resonances with optimal control, we obtain a programmable family of momentum space tight-binding models in which on-site energies are set by the chosen resonance while the nearest-neighbor tunnelling rates remain tunable, enabling the precise implementation of targeted effective Hamiltonians.

The paper is organized as follows. In Sec.~II, we develop the general Floquet formalism allowing the simulation of arbitrary one-dimensional tight-binding models in momentum space. Starting from the kicked rotor, we introduce the shaken rotor Hamiltonian and derive, within first-order time-dependent perturbation theory, the relation linking the modulation function to the effective tunneling amplitudes. This leads to a kernel equation whose inversion determines the class of experimentally accessible tight-binding models at quantum resonance.
In Sec.~III, we investigate analytically several relevant quantum resonances ($\hbar_{\mathrm{eff}} = 2\pi,\, 3\pi,\, 4\pi,$ and $6\pi$). For each case, we determine the structure of the effective onsite energies and derive explicit modulation functions. We show in particular that the principal resonance $\hbar_{\mathrm{eff}} = 4\pi$ naturally recovers the Bragg formalism, and that the infinite-chain limit connects our approach to kicked dynamics. The fidelity of the effective Floquet Hamiltonian is quantified as a function of tunneling amplitude and chain size. In Sec.~IV, we introduce an optimal-control framework to improve the agreement between the target tight-binding evolution and the exact Floquet dynamics. We define a multi-time fidelity and implement a gradient-based optimization procedure. This approach significantly enhances the dynamical fidelity over multiple periods and enables partial control of higher-order processes, including second-nearest-neighbor tunneling under specific resonance conditions.
In Sec.~V, we present experimental demonstrations of the method using a Bose–Einstein condensate in a dynamically modulated optical lattice. We first implement and characterize the Rice–Mele model, including quantum walks, band-splitting measurements, and the preparation of topological edge states at engineered junctions. We then demonstrate momentum-space Bloch oscillations and show how different initial states probe the underlying band structure. Finally, we realize superlattice configurations with higher spatial periodicity and investigate the role of second-nearest-neighbor tunneling.
In Sec.~VI, we discuss the limitations of the method. We analyze the constraints imposed by the finite quasi-momentum width of the atomic cloud, the limited bandwidth of the modulation apparatus, and the perturbative regime required for high-fidelity Floquet engineering. We also comment on intrinsic restrictions of the approach, such as the limited tunability of onsite energies and the nonlocal character of interactions in momentum space.

\section{General Floquet formalism for 1D tight-binding models}

In this section, we build the theoretical framework underlying our MSL simulator. 
We first recall how the kicked rotor maps onto a one-dimensional tight-binding model in momentum space, and emphasize the special role played by quantum resonances, where discrete translational invariance enables a Bloch-band description. 
We then introduce the shaken rotor, whose additional control parameters allow us to engineer both on-site energies and tunnelling amplitudes in the MSL. 
Finally, we derive a first-order design equation linking the modulation to the effective couplings, and reformulate it as a kernel equation to address the inversion problem.

\subsection{The kicked rotor}

Consider the kicked rotor Hamiltonian
\begin{equation}
\hat H(t)=\frac{\hat p^{\,2}}{2}\;+\;\mathcal{K}\,\cos\hat x\,\sum_{m\in\mathbb{Z}}\delta(t-mT),
\label{eq:Hkr}
\end{equation}
with $[\hat x,\hat p]=i$, kick period $T$, and kick area $\mathcal{K}$. Introducing dimensionless variables $\tau=t/T$, $\hat P=T\hat p$, $\hbar_{\rm eff}= T$, and $K=\mathcal{K}T$, the stroboscopic evolution from just before kick $m$ to just before kick $m{+}1$ is a kick followed by free rotation: $|\psi(\tau_m^+)\rangle=e^{-\,\frac{i}{\hbar_{\rm eff}}K\cos\hat x}|\psi(\tau_m^-)\rangle$ and $| \psi(\tau_{m+1}^-) \rangle=e^{-\,\frac{i}{\hbar_{\rm eff}}\frac{\hat P^{\,2}}{2}}|\psi(\tau_m^+) \rangle$.
The one-period Floquet operator is defined by 
$\hat U_F = \exp(-i\hat{H}_F/\hbar_{\rm eff})$ where $\hat{H}_F$ is the Floquet Hamiltonian. In the momentum eigenbasis restricted to the zero quasi-momentum subspace defined as $\beta_p = 0$, $\hat P|n\rangle=n\,\hbar_{\rm eff}|n\rangle$, the kick couples nearest neighbors in momentum space and the Floquet Hamiltonian can be expressed in the general form
\begin{equation}
\begin{split} 
\hat{H}_F \;=\;&
\sum_{n}\varepsilon_n\,|n\rangle\!\langle n |
\;+\;\sum_{n}\left(t^{(1)}_n\,|n+1\rangle\!\langle n|+{\rm h.c.}\right)\\
 &\;+\;\sum_{n}\left(t^{(2)}_n\,|n+2\rangle\!\langle n|+{\rm h.c.}\right)\;+\;...  ,
\end{split} 
\label{eq:HF-nbasis}
\end{equation}
The Floquet Hamiltonian corresponds to a matrix $H_{n,m} = \langle n|H_F|m\rangle$ whose diagonal elements can be considered as on-site energies $H_{n,n} = \varepsilon_n$ and off-diagonal elements as tunnelling coefficients to the first nearest neighbor $H_{n+1,n} = t^{(1)}_n$, to the second nearest neighbor $H_{n+2,n} =t^{(2)}_n$,... which generally depend on the momentum state $n$. As this paper mostly deals with the first nearest neighbor tunneling coefficient, we will often omit the exponent and write $t^{(1)}_n = t_n$. 
Thus the kicked rotor realizes a 1D tight-binding model in momentum space, however the tunneling coefficients and on site energies both depend on two parameters only, $K$ and $\hbar_{\rm eff}$, and cannot easily be tuned independently. 

In this model, quantum resonances occur when the effective Planck constant satisfies
$\hbar_{\mathrm{eff}} = 4\pi p/q$, with $p$ and $q$ integers.
For these special values, the phase accumulated during the free evolution becomes commensurate with $2\pi$, which has profound consequences in momentum space. To see this more clearly, let us first switch off the lattice and set $K=0$.
The Floquet operator over one period then reduces to the free evolution and is diagonal in the momentum basis:
\begin{equation}
\langle \ell|\hat U_F|n\rangle 
= e^{-i2\pi p n^2/q}\,\delta_{n,\ell}.
\end{equation}
Because the phase factor depends on $n^2$ modulo $q$, it satisfies
\begin{equation}
\langle \ell|\hat U_F|n\rangle
=
\langle \ell+q|\hat U_F|n+q\rangle,
\end{equation}
which means that the evolution operator is periodic in momentum space with period $q$.
This periodicity allows us to reinterpret the system as an effective lattice made of identical unit cells containing $q$ sites.
Writing any momentum index as $n=\ell_n q+r_n$, with the remainder $r_n\in[0,q-1]$, we see that the phase accumulated during one period depends only on $r_n$.
One can therefore assign to each site inside the unit cell an effective on-site quasi-energy
\begin{equation}
\epsilon_{n} = \frac{\hbar^2_{\rm eff}}{2}  r_n^2 
\quad \mathrm{mod}\; 2\pi\hbar_{\rm eff},
\label{onsiteeffenergy}
\end{equation}
where the modulo reflects the fact that quasi-energies are defined up to multiples of $2\pi$.

When the kicks are restored ($K\neq0$), tunnelling between momentum states is reintroduced, but the discrete translation symmetry $n\rightarrow n+q$ survives.
The Floquet operator can then be diagonalized using Bloch theory, exactly as for a particle moving in a spatially periodic lattice \cite{IzrailShepe_THEORMath_1980}. 
At quantum resonance, the discrete translational invariance in momentum space implies that Floquet eigenstates can be labeled by a quasi-position $\beta_x$, in full analogy with the quasi-momentum $\beta_p$ of a particle in a spatial lattice.
Physically, this quantity is conjugate to momentum and therefore corresponds to the real position.
The quasi-energy spectrum then organizes into $q$ Bloch bands as a function of $\beta_x$.
An initial state localized on a single momentum site thus spreads through a coherent quantum walk, leading to a ballistic expansion of the wave packet \cite{Ryu_PRL_2006}.
Away from quantum resonances, the translational symmetry is lost.
The phases accumulated during free evolution vary quasi-randomly from site to site, so that the on-site energies $\varepsilon_n$ effectively behave as pseudo-random variables.
The system then realizes a momentum-space version of the Lloyd--Anderson model and exhibits dynamical localization \cite{Grempel_PRA_1984,Moore_PRL_1995}.

In this article, our goal is to use momentum-space lattices as quantum simulators of tight-binding Hamiltonians with parameters that are as tunable as possible.
In other words, we would like to engineer independently the diagonal and off-diagonal matrix elements $H_{n,m}$ so that they reproduce a desired effective lattice model.
This requirement naturally leads us to operate at quantum resonance, in order to retain translational invariance, while introducing additional degrees of freedom in the time dependence of the driving beyond the single Dirac comb of the standard kicked rotor.

\subsection{The Shaken Rotor Hamiltonian}

To gain additional control over the effective lattice parameters, we consider the shaken rotor model, a generalization of the kicked rotor.
Physically, it corresponds to an atom confined in an optical lattice potential whose position and amplitude are periodically modulated in time.
The corresponding Hamiltonian reads
\begin{equation}
\hat H(t)=\frac{\hat p^{\,2}}{2m}\;-s_0(t)\frac{E_L}{2}\cos(k_L \hat x-\phi(t)),
\label{eq:Hphys}
\end{equation}
with the lattice wavevector $k_L=2\pi/d$ and energy $E_L = \hbar^2k_L^2/2m$, where $d$ is the lattice spacing, $m$ the atomic mass, and $\hbar$ Planck's constant.
The amplitude $s_0(t)$ (lattice depth) and phase $\phi(t)$ (lattice position) are the control functions, and are periodic with period $T$. 
This Hamiltonian has recently served as a framework for studying coherent scattering of matter waves in classically chaotic systems with specific symmetry properties~\cite{arrouas2023floquetoperatorengineeringquantum}. 
Our objective is to define a systematic approach to engineer the control functions $s_0(t)$ and $\phi(t)$ in order to realize physically relevant tight-binding Hamiltonians featuring periodic on-site potentials and tunable complex tunneling amplitudes between first, and possibly second, neighbors.

To establish the formalism, we introduce the following dimensionless variables: $\tilde{x} = k_L x$ and $\tilde{t} = t/T$, 
so that the wave function $\psi(\tilde{x},\tilde{t})$ satisfies the dimensionless Schrödinger equation
\begin{equation}
    i\hbar_{\mathrm{eff}} \frac{ \partial \psi}{\partial \tilde{t}} 
    = -\frac{\hbar_{\mathrm{eff}}^2}{2}\frac{\partial^2\psi}{\partial \tilde{x}^2} 
    -\frac{\tilde{s}_0}{2}\cos\!\left(\tilde{x}-\tilde{\phi}\right)\tilde{\psi},
\end{equation}
with the effective Planck constant defined as 
\(\hbar_{\mathrm{eff}} = 2E_L T/\hbar = \hbar k_L^2 T / m\), 
and the controls $\tilde{s}_0(\tilde{t}) = s_0(T\tilde{t})\,\hbar^2_{\mathrm{eff}}/2$ and $\tilde{\phi}(\tilde{t})=\phi(T\tilde{t})$.
Since the Hamiltonian is $2\pi$-periodic in position, Bloch's theorem guarantees that the quasi-momentum is conserved during the time evolution.
We therefore focus on the subspace of quasi-momentum $\beta_p = 0$ and assume that the atomic state remains confined to this sector throughout the dynamics.
Experimentally, this condition is well approximated for a Bose--Einstein condensate of weakly interacting atoms delocalized over many lattice sites.
Within this subspace, the wave function is periodic in position and can be expanded onto the plane-wave basis
$\langle \tilde{x}|n\rangle = e^{in\tilde{x}}/\sqrt{2\pi}$.
In this representation, the Hamiltonian takes the form
\begin{eqnarray}\label{Hamiltonien_q0}
    \hat{H}(\tilde{t}) & = &
    \sum_{n\in\mathbb{Z}}
    \frac{\hbar_{\rm eff}^2 n^2}{2}|n\rangle\langle n|\nonumber \\
    &+&   \sum_{n\in\mathbb{Z}} \left[\frac{f(\tilde{t})}{2}|n+1\rangle\langle n|
    + {\rm h.c.} \right]
\end{eqnarray}
\noindent where the complex modulation function is
$f(\tilde{t}) = -\tilde {s}_0(\tilde{t}) e^{-i\tilde{\phi}(\tilde{t})}/2
= f_1(\tilde{t}) + i f_2(\tilde{t})$.
Our objective is now to design a modulation $f(\tilde{t})$ capable of producing a Floquet Hamiltonian $\hat{H}_F$ with arbitrary tunnelling amplitudes between momentum sites.
As a first step, we consider a treatment based on first-order time-dependent perturbation theory.

\subsection{First order time dependent perturbation theory}

We consider arbitrary plane wave state as an initial state $|\psi(\tilde{t}=0)\rangle = |k\rangle$.
In order to disentangle the rapid phase evolution generated by the kinetic energy from the slower dynamics induced by the modulation, we factor out the free evolution and write
\begin{equation}\label{Eq_psi_t}
|\psi(\tilde{t})\rangle =
\sum_{n\in\mathbb{Z}}
e^{-iE_n \tilde{t}/\hbar_{\mathrm{eff}}}\,
\gamma_n(\tilde{t})\,|n\rangle,
\end{equation}
where $E_n = \hbar_{\mathrm{eff}}^2 n^2 / 2$.
By construction, the coefficients $\gamma_n(\tilde{t})$ describe the population transfer between momentum sites in the interaction picture defined by the kinetic Hamiltonian. Assuming weak driving, we treat the modulation perturbatively and expand
$\gamma_n = \gamma_n^{(0)} + \lambda\,\gamma_n^{(1)} + \dots$,
with $\lambda = \mathrm{max}|f(t)|/\hbar_{\mathrm{eff}}^2 \ll 1$, which measures the ratio between the induced tunnelling and the energy spacing of neighboring sites.
At zeroth order, the particle remains on its initial site, $\gamma_n^{(0)} = \delta_{k,n}$.
Because the potential couples neighboring momenta, first-order processes populate only $|k{\pm}1\rangle$.
Standard time-dependent perturbation theory then yields, after one period,
\begin{equation}
i\hbar_{\mathrm{eff}}\,\gamma_{k+1}^{(1)}(1)
=
\frac{1}{2}\int_0^1
e^{-i(E_k - E_{k+1})t'/\hbar_{\mathrm{eff}}}\,
f(t')\,d t'.
\label{pert_amp}
\end{equation}
This is the transition amplitude obtained from the exact microscopic dynamics, expressed in the interaction frame of the kinetic energies.

We now turn to the realization of an effective lattice model. Choosing $T$ in order to be at a quantum resonance, we introduce an effective Floquet Hamiltonian that we want the stroboscopic evolution to reproduce.
Restricting to first nearest neighbors, we write the Floquet Hamiltonian as
\begin{equation}
\hat{H}_F =
\sum_{n} \epsilon_n |n\rangle\langle n|
+ \sum_{n} \left[ t_n^{(1)} |n{+}1\rangle\langle n|
+ \mathrm{h.c.} \right]
\end{equation}
where the on-site energies $\epsilon_n$ are given by the quantum resonance condition in Eq.~\eqref{onsiteeffenergy}. If we consider the same initial state $|\psi(\tilde{t}=0)\rangle = |k\rangle$, we can decompose the wave function like in Eq.~(\ref{Eq_psi_t}),
\begin{equation}\label{Eq_psi_t_eff}
|\psi(\tilde{t}=1)\rangle =
\sum_{n\in\mathbb{Z}}
e^{-i\epsilon_n/\hbar_{\mathrm{eff}}}\,
\gamma_n(1)\,|n\rangle\;.
\end{equation}
We expect the tunnelling amplitudes to be much weaker than the on-site energies, which allows us to perform the same expansion of $\gamma_n(1)$ as before. At zeroth order we get $\gamma_n^{(0)} = \delta_{k,n}$, and at first order, 
\begin{equation}
i\hbar_{\mathrm{eff}}\,\gamma_{k+1}^{(1)}(1)
=
\int_0^1
e^{-i(\epsilon_k - \epsilon_{k+1})t'/\hbar_{\mathrm{eff}}}\,
t^{(1)}_k\,d t'.
\end{equation}
with a similar equation for $k-1$.
We introduce the factor 
\begin{equation}
\Gamma_k =
\int_0^1
e^{-i(\epsilon_k - \epsilon_{k+1})t'/\hbar_{\mathrm{eff}}}\,dt'\;,
\label{gamma_eq}
\end{equation}
and now equate the solutions found from the effective Hamiltonian and the time-dependent cosine potential.
Using the resonance condition to rewrite $(E_k-E_{k+1})/\hbar_{\mathrm{eff}} = -2\pi (p/q)(2k+1)$, we finally obtain
\begin{equation}
t_k^{(1)} =
\frac{1}{2\Gamma_k}
\int_0^1
e^{-i2\pi \frac{p}{q}(2k+1)t'}\,f(t')\,d t'.
\label{master_eq}
\end{equation}
Equation~(\ref{master_eq}) is at the core of our perturbative approach, as it provides a direct mapping between the experimentally controlled modulation and the nearest-neighbor tunnelling amplitudes of the effective tight-binding lattice.

We now wish to invert Eq.~(\ref{master_eq}) in order to determine a modulation $f(t)$ that reproduces a desired set of tunnelling amplitudes, given the on-site energy pattern imposed by the resonance.
A central question is therefore whether such a function exists and, if it does, whether it is unique.
Equation~(\ref{master_eq}) has the structure of a projection.
Indeed, the quantities $t_k^{(1)} \Gamma_k$ are obtained by integrating $f(t)$ against the family of phase factors
$\{ e^{4i\pi p k t/q} \}_{k\in\mathbb{Z}}$,
which form a set functions in the space of periodic signals.
In other words, determining $f(t)$ from the tunnelling coefficients amounts to reconstructing a function from its overlaps with this family.
The possibility and uniqueness of the reconstruction thus depend on how these functions span the space:
if they form a complete and linearly independent set, the inversion is well defined; otherwise, multiple solutions or no solution may exist.
In principle, higher-order perturbative processes would provide access to longer-range tunnelings such as $t_k^{(2)}$.
However, already at second order the structure of the equations becomes too involved to allow for a practical inversion.
We now reformulate Eq.~(\ref{master_eq}) as a kernel equation, which will provide a more convenient framework to design experimentally feasible solutions.

\subsection{Kernel equation for the modulation function}
Certain choices of the integers $p$ and $q$ entering the definition of $\hbar_{\mathrm{eff}}$ lead to particularly transparent analytical structures.
For instance, when $q = 2$ and $p = 1$, the family of functions
$\{ e^{4 i \pi p k t / q} \}_{k \in \mathbb{Z}}$
reduces to
$\{ e^{i 2 \pi k t} \}_{k \in \mathbb{Z}}$,
which forms the usual orthonormal Fourier basis of periodic functions.
In that situation, Eq.~(\ref{master_eq}) simply states that the tunnelling amplitudes are given by the Fourier components of $f(t)$, and the inversion becomes straightforward.
For more general values of $p$ and $q$, the situation is more involved because the corresponding family of phase factors is not orthogonal.
To make progress, it is convenient to reorganize the information contained in the set of coefficients $\{t_k^{(1)}\}$ into a generating function.
We therefore introduce the auxiliary function
\begin{equation}
    T(z) = \sum_{k\in\mathbb{Z}} t_k^{(1)} \Gamma_k\, e^{i2\pi k z}.
\end{equation}
This object can be viewed as the Fourier series whose coefficients are precisely the projected quantities entering Eq.~(\ref{master_eq}). We show in Appendix~(\ref{annexe1}) using Eq.~(\ref{master_eq}) that $T(z)$ can be written as
\begin{align*}
T(z)
&=
\int_0^1
\left[
\sum_{k\in\mathbb{Z}}
\frac{1}{2}
e^{-i2\pi \frac{p}{q}(2k+1)t' + i2\pi k z}
\right]
f(t')\,d t' \\
&= \int_0^1 K(z,t')\,f(t')\,d t',
\end{align*}
where the kernel is
\begin{equation}\label{kernel}
    K(z,t)=\frac{q e^{-i\pi z}}{4p}
    \sum_{\ell\in\mathbb{Z}}(-1)^\ell
    \delta\!\left( t-\frac{q}{2p}(z-\ell)\right).
\end{equation}
Because $K$ is a sum of Dirac-$\delta$ functions, the integral can be evaluated explicitly.
The function $T$ is therefore obtained as a superposition of shifted copies of the modulation $f$.
To make this structure more apparent, we introduce a rescaling of the variable $z\rightarrow q z/2p$, 
and define
$\tilde{T}(z) = (4p/q)e^{i2\pi p z/q}T(2p z/q)$.
We then arrive at
\begin{equation}
\tilde{T}(z)=
\sum_{\ell\in\mathbb{Z}}
(-1)^\ell\,
\mathbb{1}_{[0,1]}\!\left(z-\frac{q\ell}{2p}\right)
f\!\left(z-\frac{q\ell}{2p}\right),
\label{eq_fond}
\end{equation}
where $\mathbb{1}_{[0,1]}$ denotes the indicator function of the interval $[0,1]$.
This expression shows that $\tilde{T}$ is built from truncated (gated) replicas of $f$ translated by multiples of $q/2p$.
Because of the indicator functions, only a finite number of terms contribute for any given $z$, and the periodicity of $f$ further simplifies the reconstruction.
We stress that the kernel formulation is derived from Eq.~(\ref{master_eq}) but is not strictly equivalent to it.
The reason is that the family $\{e^{-i2\pi \frac{p}{q}(2k+1)t}\}_k$ is not orthogonal, so different modulation functions may lead to identical sets of tunnelling amplitudes.


\section{Analytical analysis of different resonances}

In this section, we solve the kernel equation for several resonance conditions
($\hbar_{\mathrm{eff}} = 2\pi$, $3\pi$, $4\pi$ and $6\pi$).
For each choice of $\hbar_{\mathrm{eff}}$, the on-site energies are fixed by the resonance and acquire a specific periodicity in momentum space.
This structure plays a central role in determining which effective tight-binding models can be engineered.
Within the present framework, only the nearest-neighbor tunnelling amplitudes $t_k^{(1)}$ can, in general, be tuned independently (including their complex phases).
To quantitatively assess the accuracy of the simulator, we compare the Floquet propagator after one period, $\hat{U}_F$, with the evolution operator $\hat{U}_T$ of the target continuous-time Hamiltonian.
We use as figure of merit the operator fidelity
\begin{equation}\label{fidelity}
    \mathcal{F} =
    \frac{1}{N^2}
    \left|
    \mathrm{tr}\!\left(\hat{U}_T^{\dagger}\hat{U}_F\right)
    \right|^2,
\end{equation}
where $N$ is the Hilbert-space dimension.

\subsection{Resonances $\hbar_{\mathrm{eff}} = 2\pi$ and $6\pi$}

For these two resonances, the denominator is $q = 2$, while the numerators are $p = 1$ or $p = 3$.
The effective on-site phase accumulated over one period is therefore
$e^{i p \pi n^{2}} = e^{i \epsilon_n / \hbar_{\mathrm{eff}}}$,
from which one finds in both cases
$\epsilon_{2n}= 0$ and 
$\epsilon_{2n+1} / \hbar_{\mathrm{eff}} = \pi$ $\mathrm{mod}\; 2\pi$.
Hence the lattice acquires a two-site periodicity: even and odd sites experience different phases.
The corresponding energy offset between the two sublattices is
$\Delta = \pi\hbar_{\rm eff}$.
This naturally enables the emulation of models with staggered on-site energies, such as the Rice--Mele model.
According to Eq.~(\ref{eq_fond}), the kernel equation simplifies considerably.
For $\hbar_{\mathrm{eff}} = 2\pi$, it becomes
\begin{equation}\label{master_eq_2pi}
	2e^{i\pi z}T(z) =
	\sum_{\ell\in\mathbb{Z}}
	(-1)^\ell\,
	\mathbb{1}_{[0,1]}(z-\ell)\,
	f(z-\ell),
\end{equation}
for $z\in\left[0,1\right]$. Because the indicator function then selects a single interval, this relation can be inverted explicitly.
One obtains for this resonance the modulation function $f_{2\pi}$ as:
\begin{eqnarray}\label{f_2pi}
	f_{2\pi}(\tilde{t})
	& = &
	2e^{i\pi \tilde{t}}\,T(\tilde{t})
	\nonumber \\
	& = &
	\frac{4i}{\pi}
	e^{i\pi \tilde{t}}
	\sum_{n\in\mathbb{Z}}
	(-1)^n t_n^{(1)} e^{i2\pi n \tilde{t}},
\end{eqnarray}

\begin{figure}[!ht]
	\centering
    \includegraphics[width=1\linewidth]{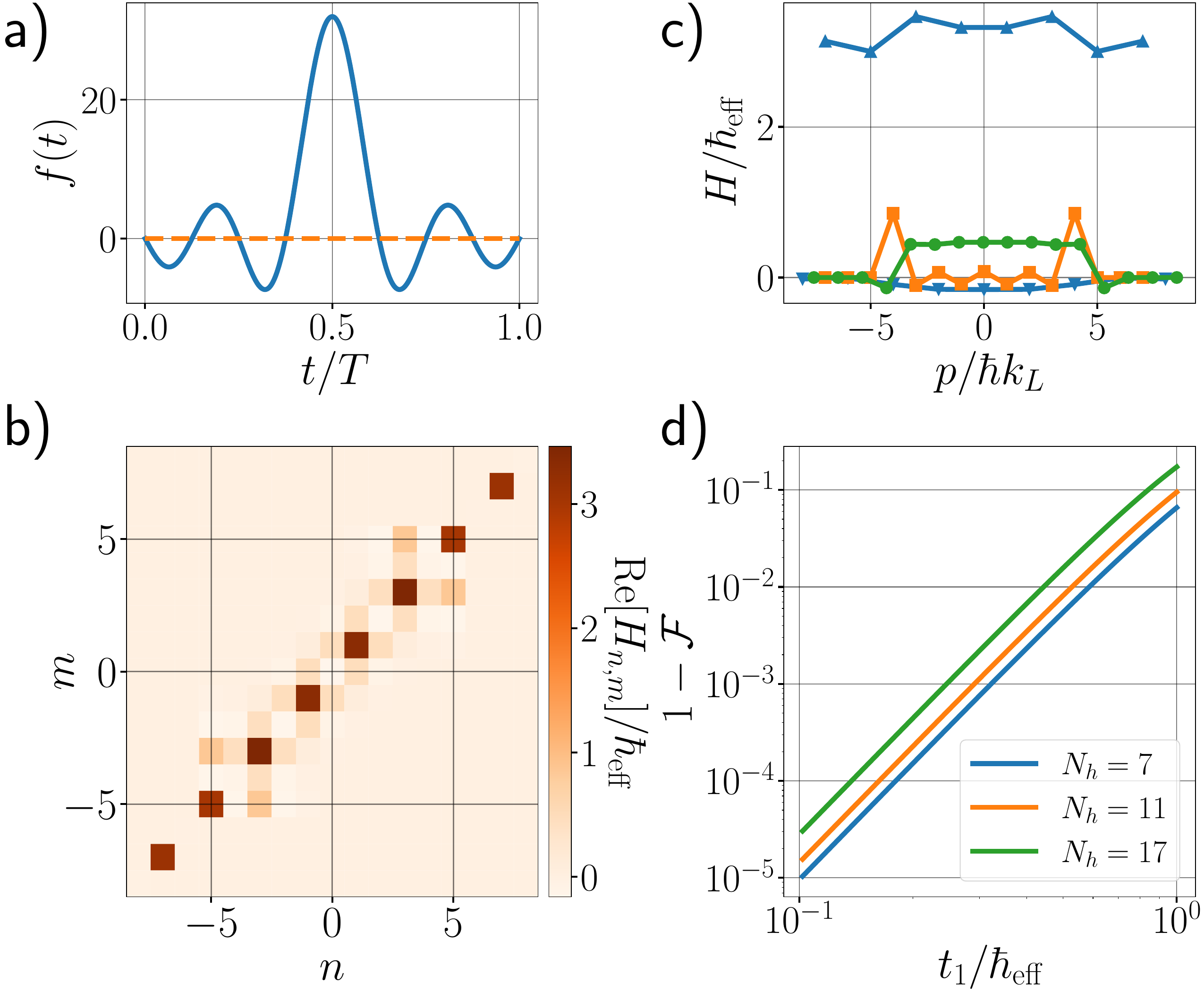}
	\caption{
	(a) Real (solid line) and imaginary (dotted line) parts of the modulation function $f(t)$ used to simulate a seven-site Rice–Mele model with a uniform tunneling amplitude $t_0 = 0.7\,\hbar_{\mathrm{eff}}$, at a $\hbar_{\rm eff}=2\pi$ resonance.  
(b) Real part of the effective Hamiltonian, $\mathrm{Re}[H_{n,m}/\hbar_{\mathrm{eff}}]$, associated with the modulation in panel~(a).  
(c) Effective Hamiltonian parameters: on-site energies $H_{n,n}/\hbar_{\rm eff}$ for even (down blue triangles) and odd (up blue triangles) sites $n$, and first- ($H_{n,n+1}/\hbar_{\rm eff}$, green circles) and second- ($H_{n,n+2}/\hbar_{\rm eff}$, orange squares) nearest-neighbor tunneling amplitudes.
(d) Fidelity between the Floquet operator generated from Eq.~(\ref{master_eq_2pi}) and the target Rice–Mele time-evolution operator, shown as a function of the target tunneling amplitude for different chain sizes. 
}
	\label{Figure_method_perturbation}
\end{figure}

To emulate a finite tight-binding chain of size $2N_h+1$, one may therefore use the modulation function given by Eq.~(\ref{f_2pi}) and impose open boundaries by setting
$t_n^{(1)} = 0$ for $|n| > N_h$. An example is shown in Fig.~(\ref{Figure_method_perturbation})a), where we choose a uniform amplitude
$t_n^{(1)} = t_0 = 0.7\,\hbar_{\mathrm{eff}}$
for $n\in[-N_h, N_h]$ with $N_h = 3$.
The resulting modulation remains continuous.
However, because periodic boundary conditions are enforced in time, the function necessarily develops sharp slope variations at the edges of the period. These rapid variations broaden the Fourier spectrum of the modulation.
This underscores the tediousness of a brute-force inversion of the system of equations given by \eqref{master_eq} in order to extract Fourier coefficients, which motivated the use of the kernel formalism.

Figure~(\ref{Figure_method_perturbation})b) shows the real part of the effective Hamiltonian extracted from the stroboscopic evolution over one driving period.
It is defined from the Floquet operator as
$\hat{H}_{n,m}/\hbar_{\mathrm{eff}} = i\, \langle n|\log(\hat{U}_F)|m\rangle$,
where $\hat{U}_F(\tilde{t}=1)$ denotes the propagator after one period.
By construction, this continuous-time Hamiltonian generates the same discrete-time dynamics as the Floquet operator.
The diagonal matrix elements, corresponding to the on-site energies
($E_{\mathrm{odd}}$ and $E_{\mathrm{even}}$ shown in Fig.~(\ref{Figure_method_perturbation})c)),
display a clear two-site periodicity with a energy offset
$\Delta \simeq \pi\hbar_{\rm eff}$, in agreement with the analytical prediction.
Moreover, the nearest-neighbor tunnelling amplitudes $t_n^{(1)}$
form a well-defined plateau over the programmed region,
with the expected width and magnitude.
This confirms that the desired tight-binding structure is correctly implemented with regards to nearest-neighbor couplings.
Nevertheless, the first-order perturbative construction does not control longer-range processes.
In particular, the next-nearest-neighbor couplings $t_n^{(2)}$
may become non-negligible and can occasionally exceed the target value of $t_n^{(1)}$.
These terms originate from higher-order virtual transitions neglected in the analytical design.
To quantify the accuracy of the simulation, we compute the fidelity defined in Eq.~(\ref{fidelity}) while varying both the system size and the target tunnelling amplitude $t_0$; the results are shown in Fig.~(\ref{Figure_method_perturbation})d).
As expected from perturbation theory, the fidelity improves for smaller chains and for weaker tunnelling.
The log--log representation reveals the approximate scaling
\begin{equation}
\mathcal{F} \approx 1 - r_{N_h}\, (t_1/\hbar_{\mathrm{eff}})^4,
    \end{equation}
with an exponent $4$ independent of the chain size $N_h$, whereas the prefactor $r_{N_h}$ depends only on $N_h$ and remains of order unity. Finally, the modulation function associated with the resonance
$\hbar_{\mathrm{eff}} = 6\pi$ follows directly from the $2\pi$ solution by a simple temporal rescaling, leading to
\begin{equation}\label{f_6pi}
	f_{6\pi}(\tilde{t})
	=
	\frac{12i}{\pi}
	(-1)^{\lfloor 3\tilde{t}\rfloor}
	e^{i3\pi\tilde{t}}
	\sum_{n\in\mathbb{Z}}
	(-1)^n t_n^{(1)} e^{i6\pi n \tilde{t}},
\end{equation}
where $\lfloor \tilde{t}\rfloor$ denotes the integer part of $\tilde{t}$ (see Appendix~\ref{annexe2}).

\subsection{Resonance $\hbar_{\rm eff}=4\pi$ and the Bragg formalism}

This quantum resonance, occurring for $p=q$, is usually referred to as the \emph{principal resonance}~\cite{IzrailShepe_THEORMath_1980}, since it is the first situation in which all effective on-site energies vanish.
In this regime, the free-evolution phase accumulated by a momentum state satisfies
$e^{2 i \pi n^{2}} = 1$, $n \in \mathbb{Z}$,
so that the kinetic contribution becomes completely trivial.
As a result, the kernel equation simplifies considerably.
Under this condition, Eq.~(\ref{eq_fond}) reduces to
\begin{equation}\label{master_eq_4pi}
    4\, e^{i2\pi z}\, T(2z) = f(z) - f\!\left(z+\tfrac{1}{2}\right).
\end{equation}
This relation does not determine $f$ uniquely.
Indeed, for any function 1-periodic function $g$, the transformation
$f(z)\rightarrow f(z) + g(z) + g(z+\tfrac{1}{2})$ leaves the right-hand side unchanged.
The general solution can therefore be written as
\begin{equation}\label{f_4pi}
    f(z) = 2\, e^{i2\pi z}\, T(2z) + g(z) + g\!\left(z+\tfrac{1}{2}\right),
\end{equation}
where $g$ is an arbitrary periodic function, and
$T(z) = \sum_{n\in\mathbb{Z}} t_n^{(1)} \, e^{i2\pi n z}$
encodes the target tunnelling amplitudes.

A particularly simple representative of this family is obtained by choosing $g=0$.
In that case,
\begin{equation}\label{f_4pi_fourier}
    f_{4\pi}(\tilde{t}) =
    2 \sum_{n\in\mathbb{Z}}
    t_n^{(1)}\, e^{i2\pi (2n+1)\tilde{t}},
\end{equation}
which shows explicitly that only odd Fourier harmonics are present.
This immediately provides a transparent physical interpretation:
the tunnelling between sites $|n\rangle$ and $|n+1\rangle$
is controlled by the spectral weight of the harmonic with frequency index $2n+1$.
This is precisely the mechanism underlying the Bragg-type resonances implemented experimentally by B.~Gadway \cite{Gadway_PRA_2015}.
In this sense, Bragg MSLs appear as a particular realization within the broader class of quantum resonances described here.

Restoring physical units, the odd harmonics correspond to the frequencies
$\omega_n = (2n+1)\,(2\pi/T)$.
Using $\hbar_{\mathrm{eff}} = 2 E_L T/\hbar = 4\pi$, this becomes $\omega_n = (2n+1)E_L/\hbar$.
Hence, here, the selection rules usually derived within the dipole approximation are directly encoded in the spectral structure of the drive at the resonance, without the need of an explicit two-level description of the Bragg processes.
It is important to stress, however, that this correspondence holds within the Floquet framework.
The effective tight-binding Hamiltonian reproduces the target dynamics only at stroboscopic times, \emph{i.e.}, at integer multiples of the driving period.
Within a single cycle, the actual motion generated by the modulation may deviate from the continuous tight-binding evolution.
These deviations become more pronounced as the chain size increases, as discussed in Sec.~\ref{infchain}.

\subsection{Resonance $\hbar_{\mathrm{eff}}=3\pi$}

This resonance allows one to implement a Rice--Mele–type model with an on-site structure different from those obtained at $\hbar_{\mathrm{eff}}=2\pi$ or $6\pi$.
The effective phases follow from
$e^{-i 3\pi n^{2}/2} = e^{-i\epsilon_n / \hbar_{\mathrm{eff}}}$,
which gives $\epsilon_{2n}/\hbar_{\mathrm{eff}} = 0$ and 
$\epsilon_{2n+1}/\hbar_{\mathrm{eff}} = -\pi/2$.
The two sublattices therefore experience an energy offset
$\Delta = -\pi\hbar_{\rm eff}/2$. Using Eq.~\eqref{gamma_eq}, the corresponding renormalization factors become
$\Gamma_{2n} = 2(1-i)/\pi$, while $\Gamma_{2n+1} = \Gamma_{2n}^{*}$.
As a consequence, the generating function entering the kernel approach reads
\[
T(z) = \frac{2}{\pi}
\sum_{k\in\mathbb{Z}}
\left( 1 + (-1)^{k+1} i \right)
t_k^{(1)} e^{i 2\pi k z}.
\]

In order to relate $f$ and $\tilde{T}$ through Eq.~\eqref{eq_fond}, and take advantage of the periodicity of $f$, it is convenient to split the sum on $\ell$ by classes of congruence modulo 3. Equation~\eqref{eq_fond} then takes the form:

\begin{eqnarray}
 3\, e^{i3\pi z/2}\, T\!\left( 3z/2 \right)
& = &
\alpha_0(z)\, f(z)
+ \alpha_2(z)\, f\!\left(z + \tfrac{1}{3}\right)
\nonumber \\
& + &
\alpha_1(z)\, f\!\left(z + \tfrac{2}{3}\right).
\label{eq_fond_3pi}
\end{eqnarray}
where the functions $\alpha_i$ are sums of indicator functions, given by
\begin{eqnarray}
    \alpha_0(z)  & = &  \sum_{j\in\mathbb{Z}} (-1)^{3j}\, \mathbb{1}_{[0,1]}(z + 2j), \nonumber \\
    \alpha_1(z)  & = &  -\sum_{j\in\mathbb{Z}} (-1)^{3j}\, \mathbb{1}_{[0,1]}(z + 2j + \tfrac{2}{3}),\nonumber \\
    \alpha_2(z)  & = &  \sum_{j\in\mathbb{Z}} (-1)^{3j}\, \mathbb{1}_{[0,1]}(z + 2j + \tfrac{4}{3}).
\end{eqnarray}

As was the case for the principal resonance, this equation admits many solutions.
One convenient representative is

\begin{equation}\label{f_3pi}
f_{3\pi}(\tilde{t}) =
\beta(\tilde{t})\, e^{i\frac{3\pi}{2} \tilde{t}}\,
T\!\left(\frac{3}{2} \tilde{t}\right),
\qquad \tilde{t}\in[0,1],
\end{equation}
where
\[
\beta(\tilde{t}) =
\begin{cases}
\frac{3}{2}, & \tilde{t} \in [0, \tfrac{1}{3}] \cup [\tfrac{2}{3}, 1], \\
3,           & \tilde{t} \in [\tfrac{1}{3}, \tfrac{2}{3}].
\end{cases}
\]
The derivation is detailed in Appendix~(\ref{annexe2}).

Interestingly, this solution is intrinsically discontinuous.
Nevertheless, when computing the Floquet Hamiltonian, one indeed recovers the expected structure:
nearest-neighbor tunnelings appear in the off-diagonal elements, while the diagonal terms display the correct two-site periodicity.

Although this ensures decently accurate dynamics over one period, we find that the dynamics generated by such quickly varying perturbative solutions often departs from the target behaviour after only a few kicks.
This limitation originates from strong intra-period micromotion, which becomes particularly relevant in larger systems.
In the following, we show that optimal-control techniques can significantly improve the long-term dynamical fidelity, even as the optimized solutions continue to display similar discontinuities.

\subsection{The infinite-chain limit: emergence of kicked modulations}
\label{infchain}

Kicked dynamics naturally emerge in the limit where the tight-binding chain becomes infinite. 
This can be shown explicitly at the quantum resonance 
$\hbar_{\mathrm{eff}} = 4\pi$, considering a simple chain with a uniform tunneling coefficient 
$t^{(1)}_n = t_0$. In this case, the time-dependent modulation function reads
$f(t) = 2t_0\sum_{n=-N_h}^{N_h} e^{i2\pi (2n+1)t}$.
In the limit $N_h \to +\infty$, this Fourier series converges toward a Dirac comb,
\begin{align*}
    f(t) 
    &\xrightarrow[N_h \to +\infty]{} 
    t_0 \sum_{n\in\mathbb{Z}} (-1)^n \delta\!\left(t-\frac{n}{2}\right).
\end{align*}
The smooth periodic modulation therefore becomes a sequence of instantaneous kicks occurring at half-integer times, with alternating signs. In this limit, the Floquet evolution operator over one period can be computed analytically and takes the symmetric form
\begin{align*}
    \hat{U} 
    = e^{-i\hat{p}^2/4\hbar_{\mathrm{eff}}}
      e^{i t_0 \cos(\hat{x})/\hbar_{\mathrm{eff}}}
      e^{-i\hat{p}^2/4\hbar_{\mathrm{eff}}}
      e^{-i t_0 \cos(\hat{x})/\hbar_{\mathrm{eff}}}.
\end{align*}
The dynamics of a wave packet can thus be interpreted as two successive lattice kicks, each separated by a free-evolution stage. Importantly, the Fourier representation of the modulation is not unique. By including even harmonics, one may instead choose
\begin{align*}
    f(t) 
    = 2t_0 \sum_{n=-2N_h+1}^{2N_h+1} e^{i2\pi n t}
    \xrightarrow[N_h \to +\infty]{} 
    2t_0 \sum_{n\in\mathbb{Z}} \delta(t-n),
\end{align*}
which corresponds to a sequence of identical kicks applied at integer times. This choice leads to the well-known Floquet operator of the quantum kicked rotor (quantum standard map)~\cite{Chirikov_SovSci_1981}:
$\hat{U} 
    = e^{-i\hat{p}^2/2\hbar_{\mathrm{eff}}}
      e^{i 2 t_0 \cos(\hat{x})/\hbar_{\mathrm{eff}}}.$
The kicked-rotor dynamics therefore emerge as the infinite-chain limit of the modulated tight-binding model.
At the quantum resonance $\hbar_{\mathrm{eff}} = 4\pi$, both forms of the Floquet operator above coincide with each other, and simplify to
$\hat{U} = e^{i 2 t_0 \cos(\hat{x})/\hbar_{\rm eff}}$.
In momentum space, this Floquet operator coincides exactly with that of an infinite tight-binding chain with uniform tunneling coefficient $t_0$. As a consequence, the population of momentum states evolves in a quantum walk, a feature that has been observed experimentally~\cite{Meier_PRA_2016}.

\section{\label{sec:level3} Optimal control improvement}

The non-uniqueness of the modulation function $f$ at several quantum resonances 
implies the existence of additional degrees of freedom, that provide a natural playground for optimal control strategies: 
by slightly adjusting the temporal modulation, one can improve the fidelity of the 
target Floquet operator. In this section, we present a systematic implementation of this approach. 
We show how optimal control algorithms can be used to refine the modulation function 
and quantitatively assess the resulting gain in Floquet fidelity.
Remarkably, this procedure also enables the controlled tuning of second-nearest-neighbor 
couplings, which are not predicted at first order within the
perturbative framework. This highlights the added flexibility offered by the control-based approach beyond perturbative intuition.

\subsection{Numerical method}

The search for improved modulation functions relies on a gradient-ascent algorithm 
designed to maximize a time-averaged fidelity. Indeed, we observed that for certain quantum resonances, such as 
$\hbar_{\mathrm{eff}} = 2\pi, 3\pi, 6\pi$, the perturbative approach can produce 
a Floquet operator whose single-period fidelity is close to unity, 
\emph{i.e.}, the overlap between the target operator and the perturbative 
Floquet operator over one period is very high. 
However, when the evolution is iterated over several periods, 
the resulting dynamics can deviate significantly from the target one.
This indicates that the typical single-period fidelities obtained from perturbative theory are not sufficient to explore multi-period dynamics.
To better constrain the evolution operator at longer times, we define a cost function based on the average fidelity between the Floquet operator 
and the target unitary evolution over $N_t$ periods:

\begin{equation}\label{Fid_multitime}
    \mathcal{F}^{(N_t)} = \frac{1}{N_t}\sum_{n=1}^{N_t}\mathcal{F}_n ,
\end{equation}
with
\begin{equation}
    \mathcal{F}_n = \frac{1}{N^2}\left|\mathrm{tr}\left(\hat{U}_T^{\dagger n}\hat{U}_F^{ n}(f_1,f_2)\right)\right|^2 .
\end{equation}
Here, $N$ denotes the Hilbert-space dimension, $N_t$ the number of modulation 
periods included in the averaging procedure, $\hat{U}_T$ the evolution operator 
associated with the target Hamiltonian over one period, and 
$\hat{U}_F(f_1,f_2) = \hat{U}(\tilde{t}=1)$ the Floquet operator of the driven system over one period, viewed as a functional of the real and imaginary part of the modulation function.
Maximizing this new fidelity enables us to correctly reproduce the stroboscopic dynamics over at least $N_t$ periods of modulation.
By construction, the multi-time fidelity $\mathcal{F}^{(N_t)}$ takes values in the interval $[0,1]$, 
and the closer it is to unity, the better the approximation of the target dynamics. 
However, this quantity should not be interpreted as a universal measure of dynamical accuracy: 
it depends on the chosen Hilbert-space truncation and does not directly quantify 
the fidelity between individual states evolved under the exact and effective dynamics.
The propagator $\hat{U}(\tilde{t})$ satisfies the time-dependent Schrödinger equation 
(Eq.~(\ref{eq:Hphys})), but for numerical purposes it is more convenient 
to work with the real-valued control functions $f_1$ and $f_2$:
\begin{align*}\label{Schrodinger_f1_f2}
    i\hbar_{\mathrm{eff}}\frac{\partial \hat{U}(\tilde{t})}{\partial \tilde{t}} 
    &= \left(\frac{\hat{p}^2}{2}+ f_1(\tilde{t})\cos(\hat{x}) + f_2(\tilde{t})\sin(\hat{x})\right) \hat{U}(\tilde{t})\\
    &= \hat{H}(f_1(\tilde{t}), f_2(\tilde{t})) \hat{U}(\tilde{t}) .
\end{align*}

In order to perform numerical optimization, the Floquet operator is approximated as a product of infinitesimal 
unitary evolutions using $M_T$ discretized time steps per period:
\begin{equation}\label{Unitary_discrete_time}
    \hat{U}(f_1,f_2) = \prod_{k=1}^{M_T}e^{-i\hat{H}(f_1^k, f_2^k)/M_T\hbar_{\mathrm{eff}}} .
\end{equation}
The product is ordered with increasing $k$ from right to left, and we have defined 
$f_j^k = f_j(k/M_T)$. 

In practice, we implement a line-search gradient-ascent algorithm. 
At each optimization step $\ell$, we compute the gradient of the average-time 
fidelity evaluated at the current modulation vector $f^{(\ell)}$ 
(which collects all parameters $f_j^{k, (\ell)}$). 
The updated modulation is then given by
\begin{equation}\label{grad_asc}
    f^{(\ell+1)} = f^{(\ell)} + \epsilon_{\ell}\times \frac{\partial \mathcal{F}}{\partial f}(f^{(\ell)}) ,
\end{equation}
where the step size $\epsilon_\ell$ may be adjusted during the optimization 
to accelerate convergence.
To perform the gradient-ascent method, we compute the gradient of the multi-time 
fidelity with respect to the control parameters $f_j^k$ (see Appendix~\ref{annexe3} 
for the complete derivation):
\begin{widetext}
\begin{equation}\label{grad_multifid}
    \frac{\partial \mathcal{F}}{\partial f_j^k} = \frac{2}{M_TN_t N^2\hbar_{\mathrm{eff}}}\sum_{n=1}^{N_t}\sum_{l=1}^n \Im \left\{\mathrm{tr}\left(\hat{U}_c^{\dagger^n}\hat{U}^{ ^n}(f_1,f_2)\right)^*\mathrm{tr}\left(\hat{U}_c^{\dagger^n}\hat{U}^{ ^{l-1}}(f)\hat{U}^B_{k+1}\frac{\partial \hat{H}(f_1^k,f_2^k)}{\partial f_j^k}\hat{U}^F_{k-1}\hat{U}^{ ^{n-l}}(f) \right)\right\} ,
\end{equation}
\end{widetext}
where $\hat{U}^F_j = \prod_{m=1}^{j}e^{-\frac{i}{M_T\hbar_{\mathrm{eff}}}\hat{H}_m}$ and $\hat{U}^B_j = \prod_{m=j}^{M_T}e^{-\frac{i}{M_T\hbar_{\mathrm{eff}}}\hat{H}_m}$  denote partial forward and backwards evolution operators, respectively.

The algorithm is stopped once the average fidelity reaches a prescribed threshold, 
typically $\mathcal{F}^{(N_t)}(f^{(\ell)}) > 0.999$, or after a maximum number 
of iterations (of the order of $\sim 300$ so that the fidelity is at least $\geq 0.98$).

	\subsection{Improvement of several resonances}
	
\begin{figure*}[!ht]
	\centering
	\includegraphics[width=1\linewidth]{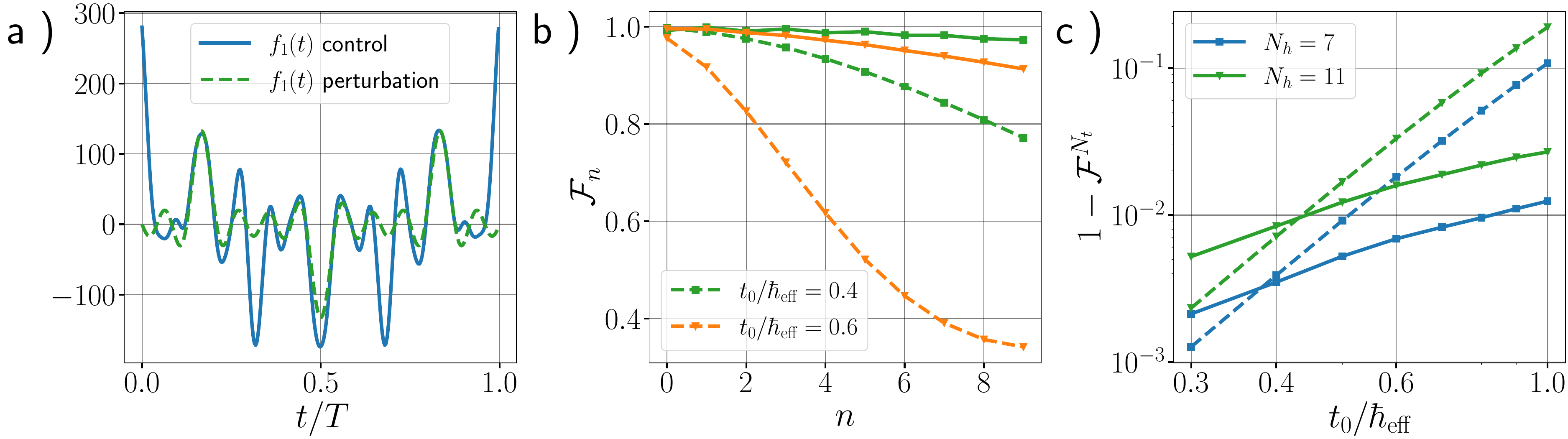}
	\caption{(a) Comparison between the modulation function obtained from first-order perturbation theory (dotted line) and its optimal-control refinement (solid line) at the quantum resonance $\hbar_{\mathrm{eff}} = 6\pi$, for a 9-site Rice–Mele model with uniform tunneling amplitude $t_0 = 0.7\,\hbar_{\mathrm{eff}}$.  
(b) Floquet-operator fidelity as a function of the number of repeated periods, for different tunneling amplitudes and $N_h=9$, comparing the perturbative solution (dotted line) and the optimal-control solution (solid line).  
(c) Time-averaged fidelity over $N_t = 5$ periods as a function of the tunneling amplitude, for chain sizes $N_h = 7$ and $11$, comparing the perturbative (dotted line) and optimal-control (solid line) approaches.
Optimal control parameters: $M_T=300$, $N=17$ and $N_t=5$.
}
	\label{Figure_controleoptimal_1}
\end{figure*}

\begin{figure}[!ht]
	\centering
	\includegraphics[width=1\linewidth]{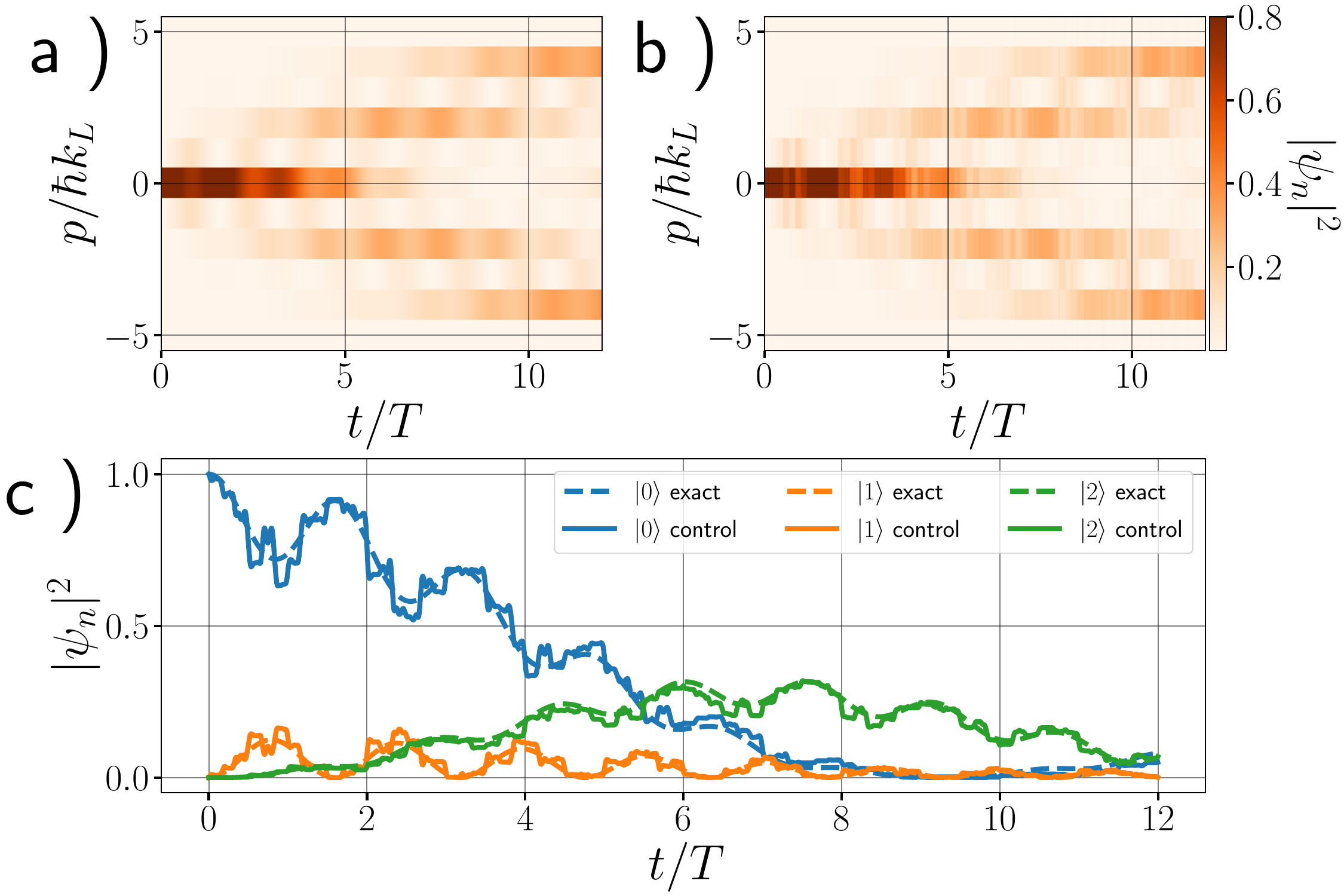}
	\caption{(a) Probability evolution map of a wave packet initially localized on site $|0\rangle$ in the exact 9-site Rice–Mele model with $\hbar_{\mathrm{eff}} = 6\pi$ and tunneling amplitude $t_0 = 0.7\,\hbar_{\mathrm{eff}}$.  
(b) Probability evolution map of the same initial state under the driven dynamics of Eq.~(\ref{Hamiltonien_q0}), using the optimal-control modulation shown in Fig.~(\ref{Figure_controleoptimal_1})a).  
(c) Time evolution of selected site populations in the exact Rice–Mele model (solid lines) compared with the optimal-control simulation (dotted lines). Optimal control parameters : $M_T=300$, $N=17$ and $N_t=5$.
}
	\label{Figure_controleoptimal_2}
\end{figure}

We used the gradient ascent method described above to refine the modulation function of a set of quantum resonances
$\hbar_{\mathrm{eff}} = 2\pi, 3\pi, 6\pi\, \text{and}\, 16\pi/3$.
The construction of the target unitary evolution proceeds as follows. 
First, we choose the periodicity and the on-site energies among the values 
allowed by the selected quantum resonance. This choice fixes 
$\hbar_{\mathrm{eff}}$ and determines the diagonal elements of the target 
Floquet Hamiltonian $\hat{H}_T$. Next, depending on the tight-binding model we wish to simulate, we assign 
non-zero tunneling coefficients to the first off-diagonal elements of 
$\hat{H}_T$. These tunneling terms are restricted to a finite subset of the 
Hilbert space, since the simulated chains are finite. All remaining matrix 
elements of $\hat{H}_T$ are set to zero. The target unitary evolution operator 
is then defined as $\hat{U}_T = \exp\left(-i\hat{H}_T/\hbar_{\mathrm{eff}}\right)$.
Starting from this definition, we apply the numerical method described above, 
using as an initial guess the modulation function obtained from the 
perturbative approach for the chosen target model. 
We emphasize that the optimization algorithm does not converge away from 
resonance, \emph{i.e.}, if the period of modulation $T$ does not verify the resonant condition $\hbar_{\rm{eff}} = \hbar k_L^2 T/m = 4\pi p/q$, nor if the diagonal elements of $\hat{H}_T$ do not match the 
on-site energies imposed by the resonance condition.
Figures~(\ref{Figure_controleoptimal_1}) and (\ref{Figure_controleoptimal_2}) 
illustrate an example for a Rice–Mele model at the quantum resonance 
$\hbar_{\mathrm{eff}} = 6\pi$, in a chain of $9$ sites with a uniform 
nearest-neighbor tunneling coefficient 
$t_n^{(1)} = t_0 = 0.7\times\hbar_{\mathrm{eff}}$ 
(for $n\in [-4,4]$ and $0$ otherwise). 
The multi-time fidelity was computed by averaging over $N_t=5$ modulation periods.

An example of the optimized modulation function is shown in 
Fig.~(\ref{Figure_controleoptimal_1})a). 
The overall shape predicted by the perturbative method is preserved, 
but additional small oscillations appear, together with sharp features 
near the beginning and the end of each period. 
These corrections cannot be anticipated from first-order perturbation theory. 
However, they can be interpreted as compensating higher-order effects, 
in particular corrections to second-nearest-neighbor tunneling amplitudes 
and on-site energies, as observed in Fig.~(\ref{Figure_method_perturbation})d).
Figure~(\ref{Figure_controleoptimal_1})b) displays the fidelity 
$\mathcal{F}_n$ as a function of the number of periods $n$. 
While the fidelity obtained with the perturbative modulation rapidly 
drops to zero, the optimized solution remains close to unity over many 
periods. This clearly demonstrates that the optimal-control modulation 
provides a much more faithful reproduction of the desired dynamics. 
As expected, the fidelity further improves for shorter chains and 
for smaller tunneling amplitudes. Moreover, the log--log plot in Fig.~(\ref{Figure_controleoptimal_2}).c) 
reveals a distinct scaling behavior of the time-averaged fidelity for 
the optimal-control solution, with a slope smaller than that obtained 
from the perturbative approach:
\[
\mathcal{F}^{N_t} \approx 1 - r_c\, \left(t_0/\hbar_{\mathrm{eff}}\right)^{\alpha}.
\]
The exponent $\alpha\approx 1.40\,(\pm0.06)$ appears to be independent 
of the chain size $N_h$, while the prefactor $r_c$ depends only on $N_h$ 
and remains of order $\sim 1$.
A numerical example of the time evolution of a wave packet initially 
localized on site $|0\rangle$ of the Rice–Mele chain is presented in 
Fig.~(\ref{Figure_controleoptimal_2})c)d)e). 
The probability dynamics is accurately reproduced, remarkably even 
within each Floquet period. 
This feature, however, is not generic: for other resonances, the 
intra-period dynamics may differ significantly, and a purely 
stroboscopic description becomes more appropriate.
Overall, the optimal-control method proves highly effective for most 
of the resonances we investigated, in particular 
$\hbar_{\mathrm{eff}} = 2\pi, 3\pi, 6\pi, 16\pi/3$.

\subsection{Controlling second nearest neighbor tunneling coefficient}

One may ask whether the optimal-control procedure could be extended 
to engineer additional tunneling amplitudes, in particular 
second-nearest-neighbor couplings. 
To address this question, we numerically explored several quantum resonances. 
We found that, in most cases, the algorithm fails to converge when such an 
additional constraint is imposed. This behavior can be understood from the structure of the modulation. 
The number of independently controllable tunneling coefficients is ultimately 
determined by the Fourier components of the control functions $f_1$ and $f_2$. 
For most resonances, these Fourier components already exhaust the available 
degrees of freedom, leaving no flexibility to independently tune 
second-nearest-neighbor tunneling amplitudes. However, we found that when $\hbar_{\mathrm{eff}} = 6\pi$, there is a narrow range of parameters where the algorithm converges. If we set the second nearest neighbor tunneling coefficient $t^{(2)} \approx 0.2 \hbar_{\mathrm{eff}}$, then it is possible to still control the first nearest coefficient $t_n^{(1)}\in [0,0.9]\hbar_{\rm{eff}}$ in the chain and obtain an average fidelity $\mathcal{F}_{N_t=5}\geq 0.98$. We can thus simulate a model with 2-periodic on site energy, fixed $t^{(2)}$ and arbitrary first tunneling coefficient, and we experimentally measured it as discusses in section \ref{Exp_Second_tunnel}.

Alternatively, a possible way to overcome this limitation would be to introduce an additional 
lattice potential with twice the spatial frequency:
\begin{align*}
    \hat{H}_2 &= g_1(t)\cos(2k_L x) + g_2(t)\sin(2k_L x)\\
    &= \sum_n g(t)|n+2\rangle\langle n| + g^*(t)|n\rangle\langle n+2|,
\end{align*}
with $g(t) = g_1(t) + i g_2(t)$.
Within first-order time-dependent perturbation theory, these new control 
functions would couple directly to second-nearest-neighbor tunneling 
coefficients, thereby providing additional independent control parameters.

\section{Experimental proof of concept and study of condensed matter models}

\subsection{Experimental setup}

Our experimental setup consists of a Bose–Einstein condensate of approximately $5 \times 10^{5}$ $^{87}$Rb atoms confined in a combined magnetic and optical dipole trap. The optical lattice is formed by two counter-propagating laser beams of wavelength $1064\,\mathrm{nm}$, aligned along the dipole trap axis. 
The beam intensities are controlled by a common acousto-optic modulator (AOM), while their relative phase is independently tuned using two separate AOMs. This configuration allows independent control of both the depth $s_0$ and  position $\phi$ of the sinusoidal potential, by driving the AOMs with properly synchronized arbitrary waveform generators.
Since the atomic cloud is sufficiently dilute, atom–atom interactions can be neglected (see discussion in Sec.~\ref{Limits_exp}). The condensate is therefore well described by a single-particle wavefunction obeying the Schrödinger equation (\ref{eq:Hphys}).
Because the spatial extent of the condensate is much larger than the lattice wavelength ($\Delta x / \lambda \approx 100$), we can assume that the system occupies the quasi-momentum subspace $q = 0$. The wavefunction can thus be expanded in the plane-wave basis $|n\rangle$, and the Hamiltonian reduces to Eq.~(\ref{Hamiltonien_q0}). The population in each plane-wave state is measured after a $35\,\mathrm{ms}$ time-of-flight and the distribution is normalized to unity in our figures.

\subsection{Experimental simulation and study of the Rice-Mele model}

The Rice–Mele (RM) model \cite{Rice_Mele_1982} is a paradigmatic one-dimensional tight-binding model describing a dimerized lattice with two inequivalent sites per unit cell, usually denoted $A$ and $B$. In our implementation, this effective lattice is realized in momentum space: the $A$ and $B$ sublattices are mapped onto even and odd momentum states, respectively, namely $|2n\rangle$ and $|2n+1\rangle$, with $n \in \mathbb{Z}$.
The two sublattices are characterized by an energy offset $\Delta$ between $A$ and $B$ sites. In addition, nearest-neighbor states are coupled by alternating tunneling amplitudes. More precisely, the coupling between $|2n\rangle$ and $|2n+1\rangle$ is given by $t_0$, while the coupling between $|2n+1\rangle$ and $|2n+2\rangle$ is given by $t_1$. This staggered tunneling structure reflects the dimerized nature of the lattice and is at the root of the model’s topological properties.
The Hamiltonian can therefore be written as
\begin{align*}
    \hat{H} &= \sum_n \Big[
    \Delta \, |2n+1\rangle\langle 2n+1|
    + t_0 \, |2n+1\rangle\langle 2n| \\
    &\qquad\qquad
    + t_1 \, |2n+1\rangle\langle 2n+2|
    + \mathrm{h.c.}
    \Big].
\end{align*}
This model is known to exhibit non-trivial topological properties. In particular, its Bloch bands are characterized by a Zak phase \cite{Zak_1989} that varies continuously with the parameters $(t_0, t_1, \Delta)$. A singular point occurs at $t_0 = t_1$ and $\Delta = 0$, where the energy gap closes and the system undergoes a topological transition. In the specific case $\Delta = 0$, the RM model reduces to the SSH model \cite{SSH_1979}, which has been extensively investigated experimentally, notably in MSLs~\cite{Meier_NatCom_2016, Xie_Nature_2019}. For a finite chain, or in the presence of topological defects, the RM model supports topological edge states localized at the boundaries between regions of different topology, like the SSH model \cite{Ullmo_Montambaux_2011}.
In our experiment, following the procedure described in Part~II, we engineer a quantum resonance such that the effective dynamics acquires a periodicity two in momentum space. This condition is fulfilled for specific values of the effective Planck constant, namely $\hbar_{\mathrm{eff}} = 2\pi$, $3\pi$, or $6\pi$. Among these possibilities, we find experimentally that $\hbar_{\mathrm{eff}} = 2\pi$ and $6\pi$ are best suited to our setup.

\begin{figure}[h]
	\centering
	\includegraphics[width=1\linewidth]{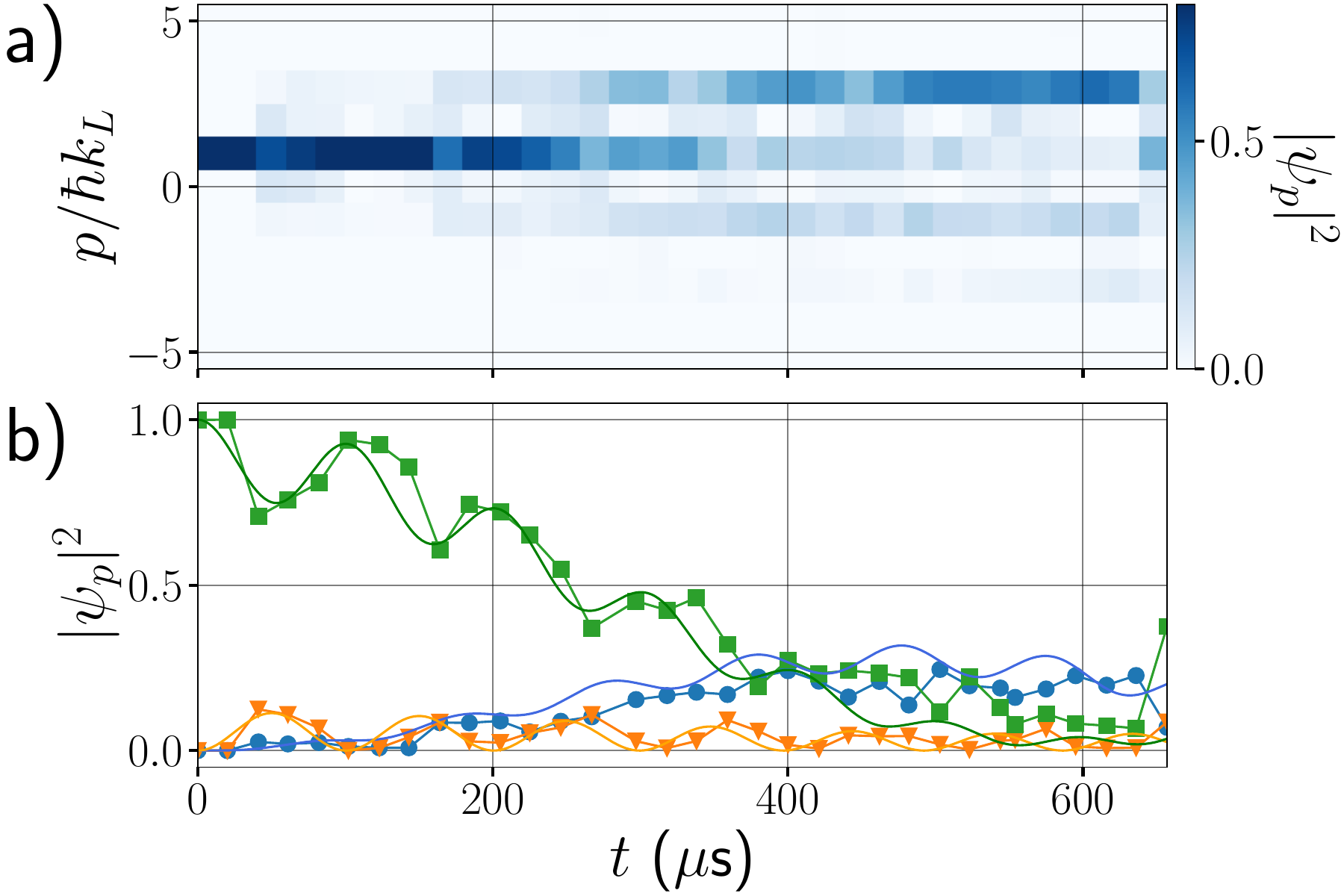}
	\caption{Experimental measurement of a quantum walk in a Rice-Mele model with $\hbar_{\mathrm{eff}}=2\pi$ so $\Delta = \pi\times\hbar_{\mathrm{eff}}$, $t_0=0.7\times\hbar_{\mathrm{eff}}=t_1$, $7$ sites and an initial state $|1\rangle$ with a modulation function improved by the optimal control method. Comparison between the time evolution of the theoretical probability  (full lines) and experimental data (markers) of the probability of being on site $|0\rangle$ (green and square dots), $|1\rangle$ (orange and round dots) and $|3\rangle$ (blue and triangular dots). The Floquet period is $T_{2\pi} \approx 61.6\,\mu$s and the optimal control parameters used to obtain the modulation function are $M_T=1000$, $N=21$ and $N_t=5$.}
	\label{Figure_Rice_Mele_Quantum_walk}
\end{figure}

In this resonant regime, the energy offset between the two sublattices is fixed by the resonance condition to $\Delta = \hbar_{\mathrm{eff}} \times \pi$. The corresponding modulation periods are $T_{2\pi} \approx 61\,\mu\mathrm{s}$ and $T_{6\pi} \approx 183\,\mu\mathrm{s}$ for $^{87}$Rb atoms. These timescales are sufficiently short to experimentally probe on the order of ten driving periods while preserving good coherence.
As a first step, we investigated a quantum walk in a finite chain of seven sites with uniform tunneling amplitudes $t_0 = t_1 = 0.7 \times \hbar_{\mathrm{eff}}$. Using an optimal control protocol \cite{Dupont_PRXQuantum}, we prepared the initial state $|1\rangle$ with high fidelity.
The dynamics shown in Fig.~(\ref{Figure_Rice_Mele_Quantum_walk}) exhibits the characteristic ballistic spreading of a quantum walk: the width of the momentum distribution increases linearly with time, with a characteristic timescale set by the tunneling amplitude. However, the presence of the staggered onsite energy $\Delta$ introduces an additional structure in the dynamics. Because energy is approximately conserved during the evolution, an initial state predominantly occupying high-energy sites tends to remain localized on high-energy sites, while a state initially prepared on low-energy sites preferentially populates other low-energy sites. As a result, the population distribution develops a clear imbalance between even and odd momentum states during the evolution. This constitutes a first experimental indication that our method effectively simulates a controllable onsite energy offset between the two sublattices. In addition to the ballistic spreading, we observe fast oscillations in the weakly populated sites, with a characteristic timescale determined by the energy offset $\Delta$. Finally, for the initial state $|1\rangle$, the quantum walk eventually reaches the boundaries of the seven-site chain and reflects from them, leading to a visible ``bounce'' at the edges.

In order to highlight the emergence of the double-band structure and to measure how the RM bandgap depends on the tunneling amplitude (with $t_0 = t_1$), we implemented the following protocol. We first prepare experimentally the ground state of a simple tight-binding chain with seven sites, in the case $\Delta = 0$ and $t_0 = t_1 = 0.7 \times \hbar_{\mathrm{eff}}$. In this configuration, the ground state reads
$|\psi_0\rangle = \mathcal{N} \sum_{n=-3}^{3} \cos\left(\frac{\pi n}{7}\right) |n\rangle$,
where $\mathcal{N}$ is a normalization constant. This state is delocalized over the entire chain and corresponds to a well-defined quasi-position $\beta_x = 0$, \emph{i.e.}, to a wave packet centered at the minimum of the effective lattice potential. We then suddenly switch on the Rice–Mele dynamics by introducing the energy offset $\Delta = \hbar_{\mathrm{eff}} \times \pi$. Because the quasi-position $\beta_x$ is conserved during the evolution, the initial state $|\psi_0\rangle$ decomposes only onto the two Rice–Mele eigenstates with the same quasi-position $\beta_x = 0$. These two eigenstates, denoted $|\psi_{\beta_x=0}^{\pm}\rangle$, belong to the upper and lower bands of the double-band structure $E^{\pm}(\beta_x)$.

\begin{figure}[!ht]
	\centering
	\includegraphics[width=1\linewidth]{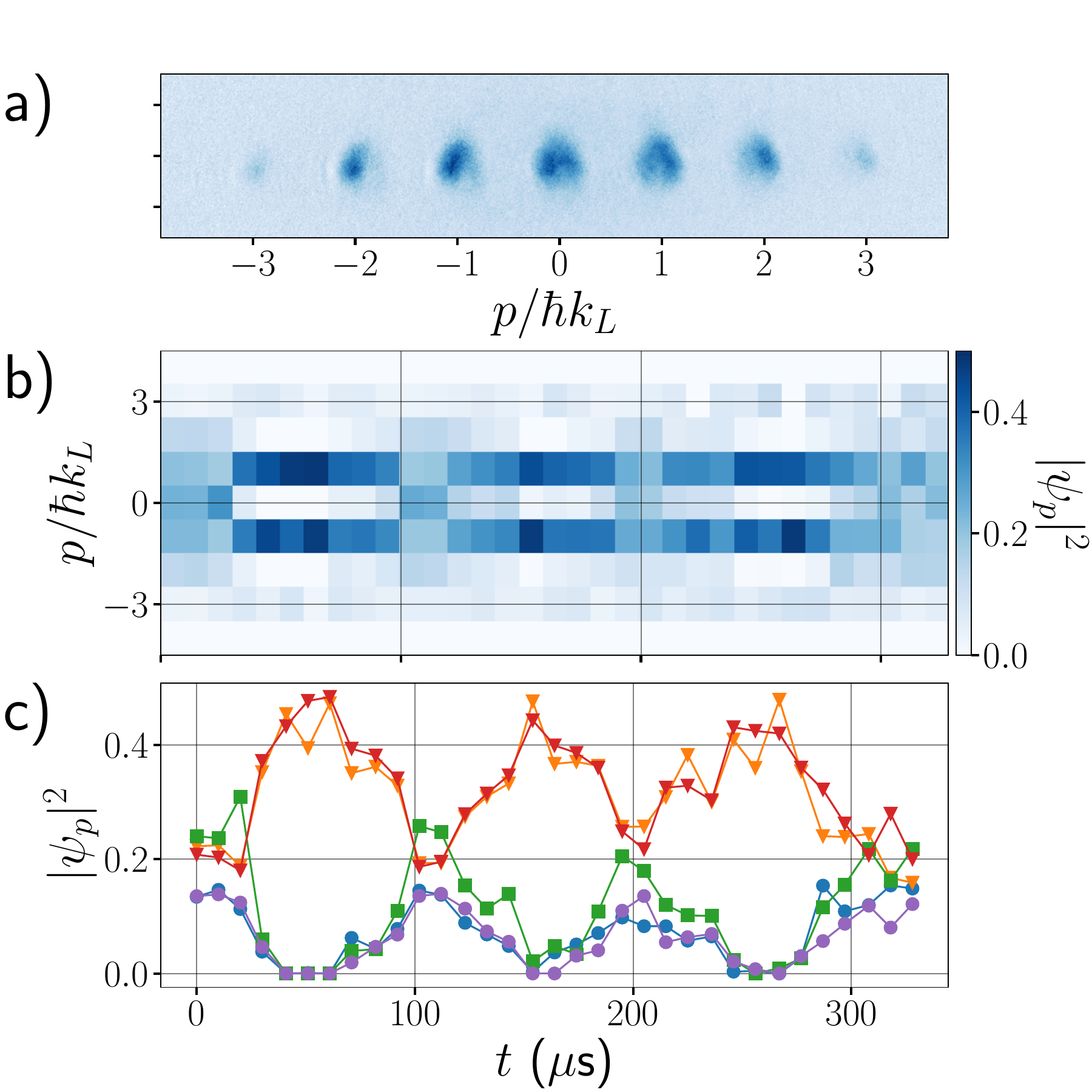}
	\caption{(a) Prepared fundamental state distribution: experimental absorption image of the momentum orders after a $35$\,ms time of flight. (b,c) Experimental observation of the coherent oscillations of the initial ground state in a seven-site Rice–Mele model at $\hbar_{\mathrm{eff}} = 2\pi$ ($\Delta = \pi\,\hbar_{\mathrm{eff}}$) with tunneling amplitude $t_0 = 0.6\,\hbar_{\mathrm{eff}}=t_1$, using an optimal-control modulation. Single-shot measurements of the site populations are shown for $|0\rangle$ (green squares), $|1\rangle$ (red triangles), $|-1\rangle$ (orange triangles), $|2\rangle$ (purple circles), and $|-2\rangle$ (blue circles). The Floquet period is $T_{2\pi} \approx 61.6\,\mu$s and the optimal control parameters used to obtain the modulation function are $M_T=1000$, $N=21$ and $N_t=5$.
}
	\label{Figure_Rice_Mele_Oscillation_Fonda}
\end{figure}

As a result, the subsequent time evolution reduces to coherent oscillations between these two states:
\begin{equation}
    |\psi(t)\rangle 
    = a\,|\psi_{\beta_x=0}^{+}\rangle 
    + b\,e^{-i\Delta_0 t/\hbar_{\mathrm{eff}}}
    |\psi_{\beta_x=0}^{-}\rangle,
\end{equation}
where $a = \langle \psi_{\beta_x=0}^{+}|\psi_0\rangle$, 
$b = \langle \psi_{\beta_x=0}^{-}|\psi_0\rangle$, 
and $\Delta_0 = E^{-}(\beta_x=0) - E^{+}(\beta_x=0)$ is the band splitting at $\beta_x=0$.
Experimentally, we indeed observe that the populations of the momentum states oscillate at a single well-defined frequency (see Fig.~(\ref{Figure_Rice_Mele_Oscillation_Fonda})a)b), confirming that only the two bands at fixed quasi-position are involved in the dynamics. In this symmetric case ($t_0 = t_1$), the oscillation frequency is directly given by the band separation at the chosen quasi-position:
\begin{equation}
\omega_{RM} 
= \frac{1}{\hbar_{\mathrm{eff}}}
\sqrt{\Delta^2 + 16 t_0^2 }.
\end{equation}
 Performing this measurement for $5$ values of the tunneling coefficient $t_0/\hbar_{\rm eff}$ between $0.1$ and $0.6$, and fitting the measured oscillation frequency, we extract the onsite energy difference between even and odd sites and obtain $\Delta^2/(\pi \hbar_{\mathrm{eff}})^2 = 0.96 \pm 0.02$, in excellent agreement with the expected value of 1.

\begin{figure}[!ht]
	\centering
	\includegraphics[width=1\linewidth]{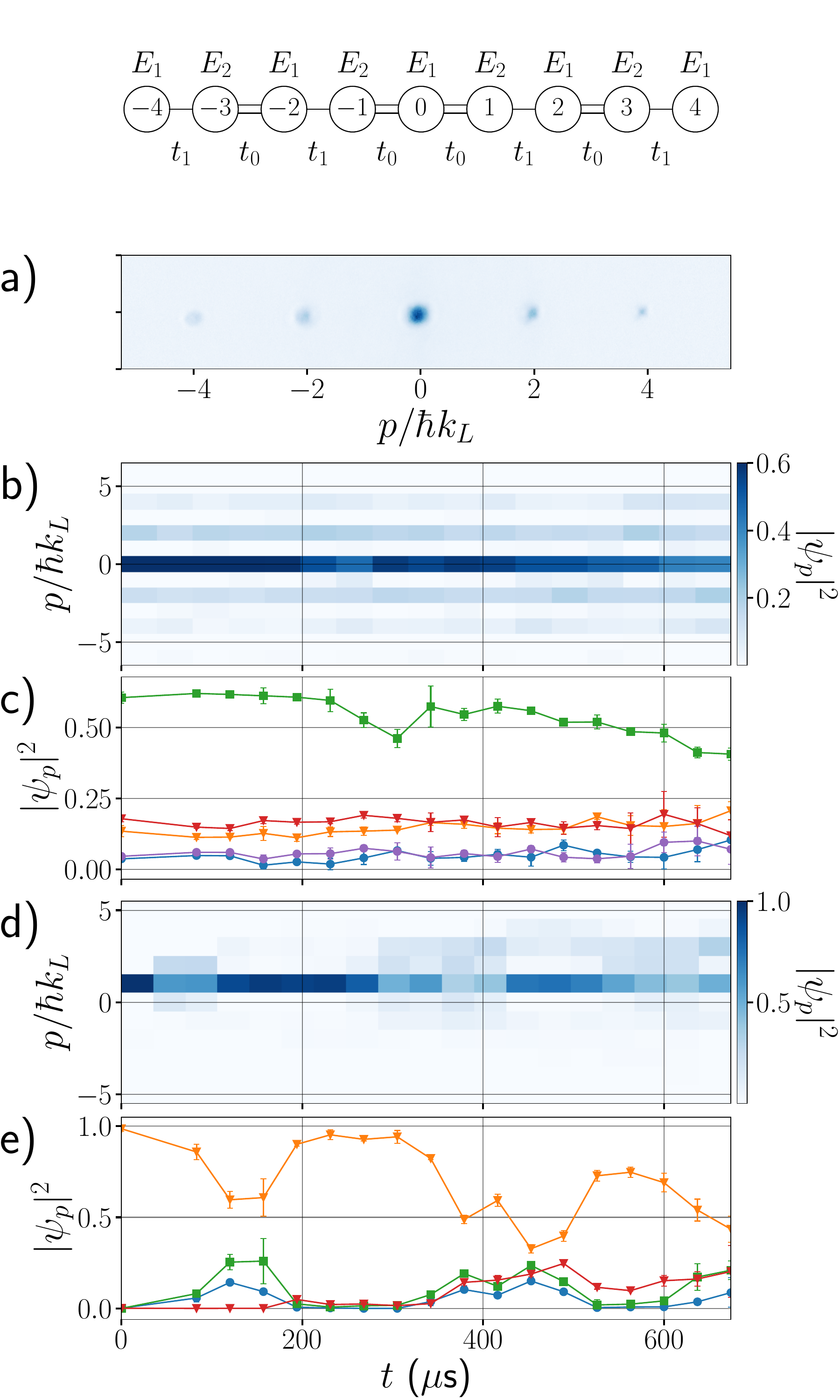}
	\caption{(a) Prepared edge state distribution: experimental absorption image of the momentum orders after a $25$\,ms time of flight. The populations on integer multiples of $\hbar k_L$ correspond to the initial populations on the MSL sketched above.
(b) Time evolution of the momentum distribution for an initial state being the  edge state exponentially localized on site $|0\rangle$, shown in a).  
(c) Time evolution of selected site populations from b): $|0\rangle$ (green squares), $|2\rangle$ (red triangles), $|-2\rangle$ (orange triangles), $|4\rangle$ (purple circles), and $|-4\rangle$ (blue circles).  
(d) Time evolution of the momentum distribution for the initial state $|1\rangle$.  
(e) Time evolution of selected site populations from d): $|1\rangle$ (orange triangles), $|2\rangle$ (green squares), $|0\rangle$ (blue circles), and $|3\rangle$ (red triangles). The Floquet period is $T_{6\pi} \approx 185.0\,\mu$s and the optimal control parameters used to obtain the modulation function are $M_T=1000$, $N=21$ and $N_t=5$.
 }
	\label{Figure_Rice_Mele_etat_de_bord}
\end{figure}

MSLs are particularly well suited for simulating defects in tight-binding models, since both the tunneling amplitudes and the chain length can be tuned independently and with high flexibility. This level of control enables the engineering of sharp interfaces and topological junctions in a fully programmable manner. In this context, we simulate a junction between two Rice–Mele chains characterized by alternating tunneling amplitudes $t_0 = 0.45 \times \hbar_{\mathrm{eff}}$ and $t_1 = 0.9 \times \hbar_{\mathrm{eff}}$, over 9 sites, operated at the quantum resonance $\hbar_{\mathrm{eff}} = 6\pi$. A defect is introduced at site $|0\rangle$, thereby creating a boundary between two regions with distinct dimerization patterns, as illustrated in Fig.~(\ref{Figure_Rice_Mele_etat_de_bord}). Using an optimal control protocol \cite{Dupont_PRXQuantum}, we experimentally prepare the topological edge state localized at the junction (Fig.~(\ref{Figure_Rice_Mele_etat_de_bord})a)), with an exponential distribution predominantly over even sites. In Figure~(\ref{Figure_Rice_Mele_etat_de_bord})b),c) we report the temporal evolution of this state, and observe that it remains stationary over time, as expected. In sharp contrast, the evolution of the initial state $|1\rangle$ prepared in the same system, Figure~(\ref{Figure_Rice_Mele_etat_de_bord})d),e) shows dynamics reminiscent of an asymmetric quantum walk, induced by the presence of the defect. This behavior can be understood from the sublattice structure of the edge mode: the state $|1\rangle$, starting on the odd sublattice, has negligible overlap with the edge state and therefore does not inherit its localization properties.

\subsection{Experimental simulation and study of the Momentum Bloch Oscillations}

Bloch oscillations are a well-known phenomenon in condensed matter physics that occurs when electrons in a crystal lattice are subjected to a weak and uniform external force. Under adiabatic conditions, a particle initially prepared with a well-defined quasi-momentum $\beta_p$ in a single energy bands remains within this band. Because the quasi-momentum increases linearly in time under the applied force according to the semiclassical equation
$\beta_p(t) = \beta_{p,0} + F t$, and the Brillouin zone is periodic, this evolution results in a periodic motion in real space. 

In cold atoms experiments, the observation of Bloch oscillations requires an initial state broadly delocalized and coherent over the lattice, often produced by loading the BEC at rest in the ground state of the lattice potential, with quasi-momentum $\beta_{p,0}=0$. Bloch oscillations are then triggered by applying a weak uniform force, usually created by a magnetic field gradient.
This behavior has been observed experimentally, either through time-of-flight 
measurements \cite{dahan_bloch_1996} or by directly tracking the center-of-mass 
motion of the atomic cloud in position space \cite{geiger_observation_2018}.

A minimal description of this phenomenon relies on a tight-binding model of a lattice with uniform nearest-neighbor tunneling amplitudes $t^{(1)}_n = t_0$ and a linearly varying onsite energy with slope $dF$, where $d$ denotes the lattice spacing and $F$ the applied force. The corresponding Hamiltonian reads
\begin{equation}\label{Bloch_Hamiltonian}
    \hat{H} = t_0 \sum_n \left( |n+1\rangle\langle n| + \mathrm{h.c.} \right)
    + dF \sum_n n \, |n\rangle\langle n|.
\end{equation}
The dynamics generated by this Hamiltonian is fully tractable and has been studied extensively \cite{hartmann_2004}. For an initial state localized on a single site, $|\psi(0)\rangle = |k\rangle$, the probability amplitude to occupy site $|n\rangle$ at time $t$ is given by
\begin{equation}\label{Solution_Bloch}
    \langle n|\psi(t)\rangle 
    = J_{n-k}\!\left( \frac{2t_0}{dF}
    \sin\left(\frac{\omega_B t}{2}\right) \right)
    e^{i(n-k)(\pi-\omega_B t)/2},
\end{equation}
where $\omega_B = dF/\hbar_{\mathrm{eff}}$ is the Bloch frequency and $J_n$ denotes the $n$th Bessel function of the first kind. As a consequence, for a generic initial state, the site populations $|\langle n|\psi(t)\rangle|^2$ evolve periodically in time with period $T_B = 2\pi/\omega_B$. Because the Bessel functions $J_{n-k}(x)$ decay rapidly for $|n-k| \gg |x|$, the wave packet remains spatially confined: its maximal spatial extension is of the order of $2t_0/dF$ lattice sites. 
Although one might naively expect the average position of the wave packet to drift in the direction of the applied force, the lattice periodicity prevents any net transport. This absence of directed motion is a direct consequence of the bounded energy band structure.

This simple tight-binding model can be mapped onto MSLs. To do so, we adopt the strategy used in \cite{Meier_PRA_2016} which can be expressed in our framework the following way.
We choose a periodic modulation with a resonance condition corresponding to 
$\hbar_{\mathrm{eff}} = 4\pi$, such that all onsite energies are effectively identical. 
The linear potential term of Eq.~(\ref{Bloch_Hamiltonian}) 
is not implemented directly as a static energy gradient, but we introduce instead a 
time-dependent phase in the tunneling amplitudes and consider the Hamiltonian
\begin{equation}
    \hat{H} = 
    t_0 e^{i\omega_B t}
    \sum_n |n+1\rangle\langle n| 
    + \mathrm{h.c.}
\end{equation}
This Hamiltonian and the tilted-lattice Hamiltonian of 
Eq.~(\ref{Bloch_Hamiltonian}) are related by the 
time-dependent unitary transformation
\begin{equation}
    \hat{U}(t) 
    = \exp\!\left(
    -i \omega_B t
    \sum_n n |n\rangle\langle n|
    \right),
\end{equation}
which corresponds to moving into an accelerated frame. 
Through this transformation, the linear potential is gauged away and replaced by a 
time-dependent Peierls phase imprinted on the tunneling coefficients. 

This mapping is particularly convenient for MSLs, 
where the tunneling phases can be directly engineered.
Experimentally, this is realized using the modulation function
\begin{equation}\label{f_4pi_Bloch}
    f(\tilde{t}) 
    = 2 \sum_{n=-N_h}^{N_h} 
    t_0\, e^{i\omega_B \tilde{t}} 
    e^{i2\pi(2n+1)\tilde{t}}.
\end{equation}
Strictly speaking, this modulation is no longer perfectly periodic due to 
the additional factor $e^{i\omega_B \tilde{t}}$. 
However, in the regime $\omega_B \ll 2\pi$, the deviation from periodicity 
is negligible on the experimental timescales considered.

\begin{figure}[!ht]
	\centering
	\includegraphics[width=1\linewidth]{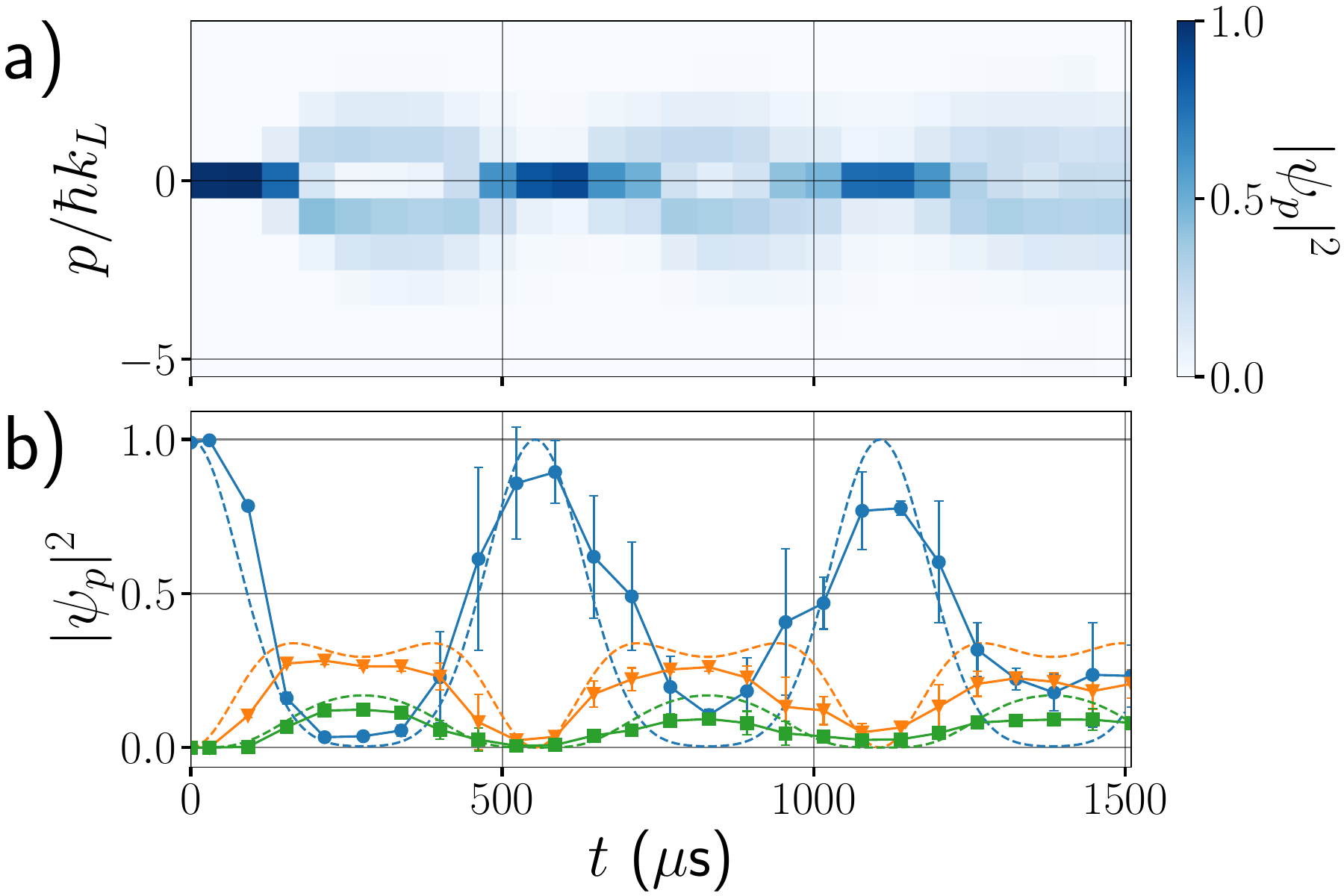}
	\caption{(a) Time evolution of the momentum distribution for an initial state localized on site $|0\rangle$ in a tight-binding model with tunneling amplitude $t_0/\hbar_{\mathrm{eff}} = 0.8$ and effective uniform force $\omega_B = 1.4$.  
(b) Time evolution of the populations of sites $|0\rangle$ (blue circles), $|1\rangle$ (orange triangles), and $|2\rangle$ (green squares), averaged over five experimental runs. The analytical prediction from Eq.~(\ref{Solution_Bloch}) is shown as dotted lines. The Floquet period is $T_{4\pi} \approx 123.3\,\mu$s.
}
	\label{Figure_Bloch_Oscillation_OD0}
\end{figure}

We experimentally measured the evolution under such a modulation of an initial state localized on site $|0\rangle$. 
This evolution in tilted MSLs was previously observed in \cite{Meier_PRA_2016}. 
The results, presented in Figure~(\ref{Figure_Bloch_Oscillation_OD0})a)b) show the spread of the initial wavepacket over approximately five sites at maximum extension, and its subsequent refocussing after four Floquet periods, clearly demonstrating coherent oscillations, that repeat at each Bloch period.

\begin{figure}[!ht]
	\centering
	\includegraphics[width=1\linewidth]{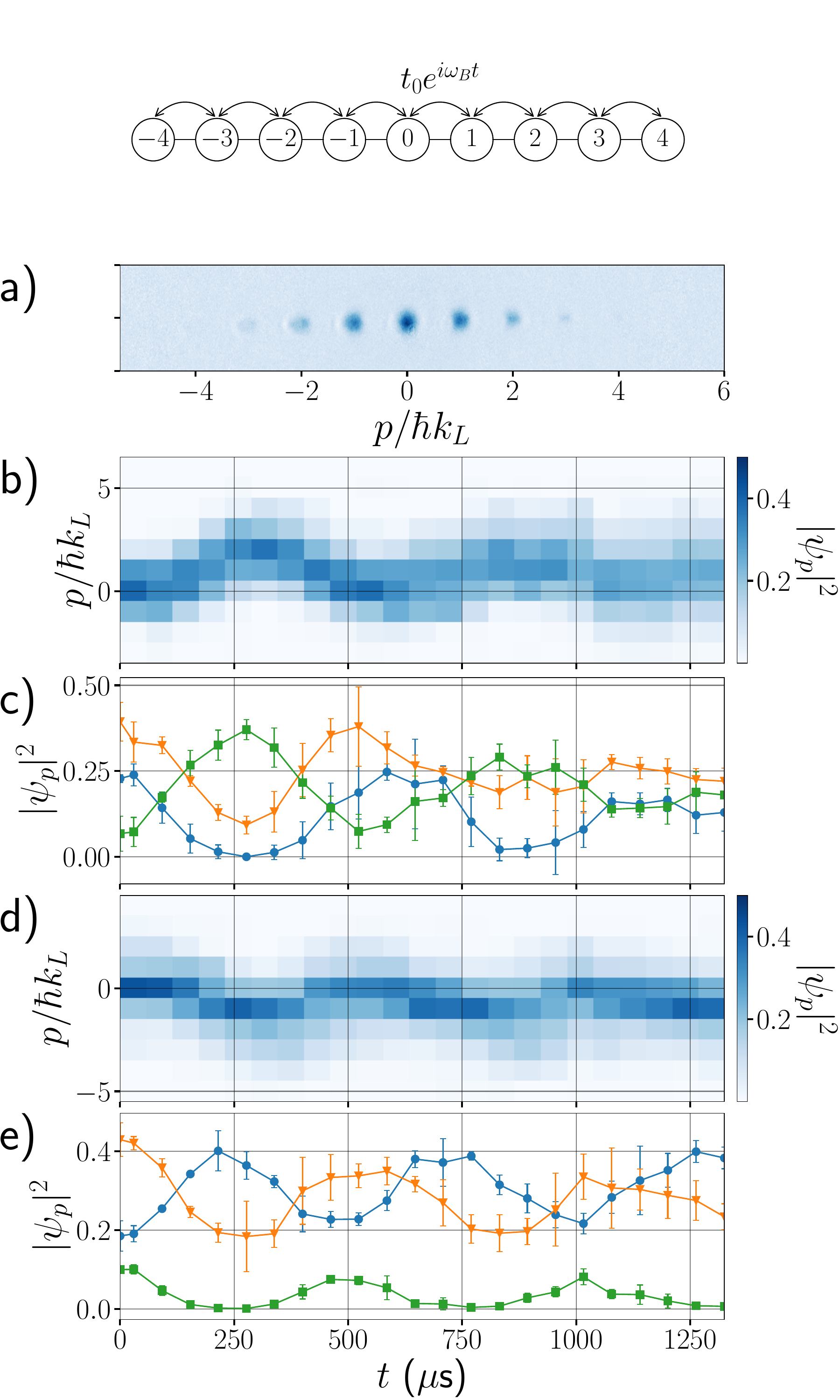}
	\caption{(a) Experimental absorption image of the diffraction orders after a $25$\,ms time of flight. The populations on integer multiples of $\hbar k_L$ correspond to the initial populations on the MSL sketched above, for a Gaussian initial state.  
(b) Time evolution of the momentum distribution from the initial Gaussian state in a tight-binding model with tunneling amplitude $t_0/\hbar_{\mathrm{eff}} = 0.8$ and effective uniform force $\omega_B = 0.7$, showing momentum space Bloch oscillations.  
(c) Time evolution of the populations of sites $|0\rangle$ (orange triangles), $|1\rangle$ (green squares), and $|-1\rangle$ (blue circles).  
(d) Time evolution of the momentum distribution from the initial alternating Gaussian state.  
(e) Time evolution of the populations of sites $|0\rangle$ (orange triangles), $|1\rangle$ (green squares), and $|-1\rangle$ (blue circles). All data are averaged over five experimental runs, the Floquet period is $T_{4\pi} \approx 123.3\,\mu$s.
  }
	\label{Figure_Bloch_Oscillation_Gaussian}
\end{figure}

The term of Bloch oscillation more commonly refers to the oscillations of wavepacket that is initially broadly spread over the lattice so that it has a relatively well-defined quasi-momentum. Translated here to the MSL framework, the roles of position and momentum are effectively interchanged compared to conventional realizations. In our configuration, recall that quasi-momentum is fixed to $\beta_p = 0$, and preparing a wave packet widely spread over momentum states corresponds to preparing a state periodic in real space and peaked
at the center of a each lattice well, \emph{i.e.}, with well-defined quasi-position $\beta_{x,0} = 0\,\mathrm{mod}\;2\pi = 0$. MSL Bloch oscillations therefore manifest as oscillations of the wave packet in momentum space, while the quasi-position evolves linearly in time: $\beta_x(\tilde{t}) = \beta_{x,0} + \omega_B\,\tilde{t}$.
To initialize the dynamics with a sufficiently delocalized state in our momentum-space lattice, we use an optimal control procedure 
\cite{Dupont_PRXQuantum} to prepare a Gaussian wave packet centered on $\ket{0}$ with width $\sigma = 2$ sites (see Fig.~(\ref{Figure_Bloch_Oscillation_Gaussian})a):
\begin{equation}
    \langle n|\psi_0\rangle \propto 
    e^{-n^2/2\sigma^2},
\end{equation}
a state which was inaccessible in \cite{Meier_PRA_2016}.
This width is large enough to yield a well-defined quasi-position and to observe the expected dynamics. 
We thus report the first experimental observation of Bloch oscillations of a wavepacket in MSLs, as shown in 
Fig.~(\ref{Figure_Bloch_Oscillation_Gaussian})b)c). Bloch oscillations appear as periodic oscillations of the average momentum $\langle \hat{p}\rangle$ of the Gaussian wave packet, while its width $\langle \Delta\hat{p}\rangle$ remains essentially unchanged. We observe this behavior over two Bloch periods.

The group velocity of the wave packet is given by the derivative of the energy band with respect to the quasi-position:
\begin{equation}
    \langle v \rangle 
    = \frac{\partial E}{\partial \beta_x}.
\end{equation}
For our lattice, the dispersion relation reads
$E(\beta_x) = -2 t_0 \cos(\beta_x)$,
and the initial Gaussian state is peaked around $\beta_x = 0$ with 
a width on the order of $1/\sigma$. 
As quasi-position increases with time, the band slope is first positive. We indeed observe that the wave packet initially drifts toward positive momenta. To further confirm this interpretation, we prepare a second initial state 
with quasi-position $\beta_{x,0} = \pi$, corresponding to a wave packet centered near the edge of the Brillouin zone. 
Experimentally, this is realized by preparing an "alternating" Gaussian state:
\begin{equation}
    \langle n|\psi_0\rangle 
    \propto (-1)^n e^{-n^2/2\sigma^2}.
\end{equation}
This initial wave packets first experiences negative group velocities. As shown in Fig.~(\ref{Figure_Bloch_Oscillation_Gaussian})d)e), we indeed observe Bloch oscillations symmetric to those of the regular Gaussian state. 
This behavior provides a direct and quantitative confirmation of the band-structure interpretation of momentum-space Bloch oscillations.

\subsection{Experimental simulation and study of the resonance $\hbar_{\mathrm{eff}} = 16\pi/3$}

\begin{figure}[!ht]
	\centering
	\includegraphics[width=1\linewidth]{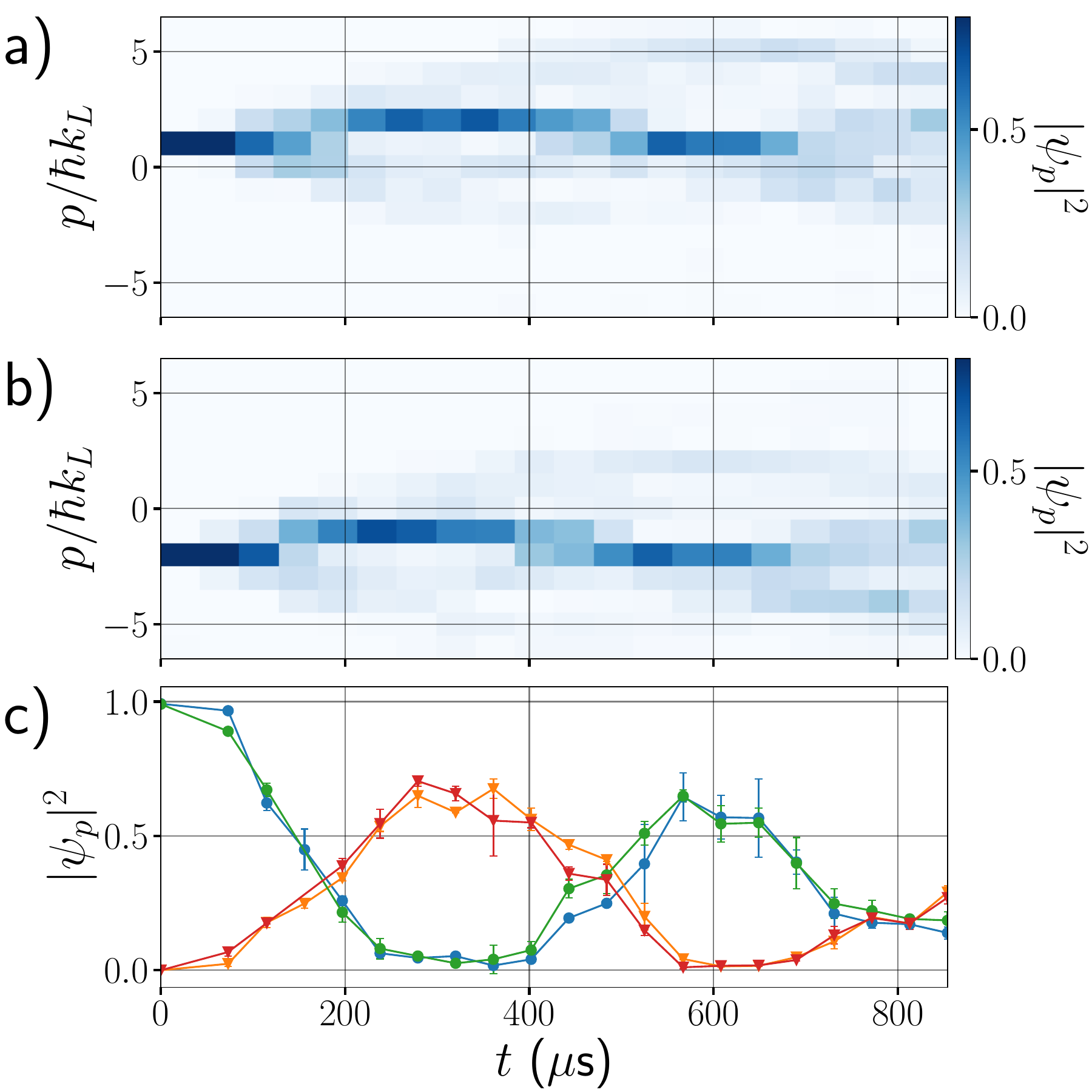}
	\caption{(a,b) Time evolution of the momentum distributions from initial states $|1\rangle$ and $|-2\rangle$, respectively, in a tight-binding model with three-site periodicity ($\hbar_{\mathrm{eff}} = 16\pi/3$) and nearest-neighbor tunneling amplitude $t_0/\hbar_{\mathrm{eff}} = 0.9$.  
(c) Time evolution of selected site populations: for the initial state $|1\rangle$, sites $|1\rangle$ (blue circles) and $|2\rangle$ (orange triangles); for the initial state $|-2\rangle$, sites $|-2\rangle$ (green squares) and $|-1\rangle$. Data are averaged over five experimental runs.  The Floquet period is $T_{16\pi/3} \approx 164.4\,\mu$s and the optimal control parameters used to obtain the modulation function are $M_T=1000$, $N=21$ and $N_t=5$.
}
	\label{Figure_Resonance_heff16pis3}
\end{figure}

In this section, we demonstrate experimentally that our method allows the simulation of tight-binding models beyond the Rice–Mele configuration. As a concrete example, we implement a lattice with a three-site periodicity. To engineer such a structure, we operate at the quantum resonance $\hbar_{\mathrm{eff}} = 16\pi/3$. For this value of the effective Planck constant, the onsite energies acquire a periodic pattern with period three: $\varepsilon_{3n} = 0$, $\varepsilon_{3n+1}/\hbar_{\mathrm{eff}} = \varepsilon_{3n+2} /\hbar_{\mathrm{eff}}= 2\pi/3$.
We numerically determine the corresponding modulation functions that realize a nearest-neighbor tight-binding model with uniform tunneling amplitude $t^{(1)} = t_0 = 0.9 \times \hbar_{\mathrm{eff}}$ on a lattice of 13 sites (corresponding to approximately four spatial periods). To verify experimentally that the engineered model indeed exhibits a three-site periodicity, we measure the evolution from two initial wave packets localized on the states $|-2\rangle$ and $|1\rangle$. 
These two initial sites are separated by exactly one spatial period of the lattice: we therefore expect their subsequent dynamics to be identical up to a translation by three sites.
The measured evolutions, shown in Fig.~(\ref{Figure_Resonance_heff16pis3}), confirm this expectation, and provide direct evidence of the imposed three-site periodic structure. Moreover, the presence of the onsite energy offset between the sites $|3n\rangle$ and the higher-energy sites $|3n+1\rangle$ and $|3n+2\rangle$ strongly influences the transport properties. Because of this energy imbalance, atoms are less likely to occupy the $|3n\rangle$ sites during the evolution, resulting in an asymmetric quantum walk. This asymmetry directly reflects the engineered three-site superlattice structure.

\subsection{Experimental control of the second nearest neighbor tunneling coefficient}\label{Exp_Second_tunnel}

\begin{figure*}[!ht]
	\centering
	\includegraphics[width=1\linewidth]{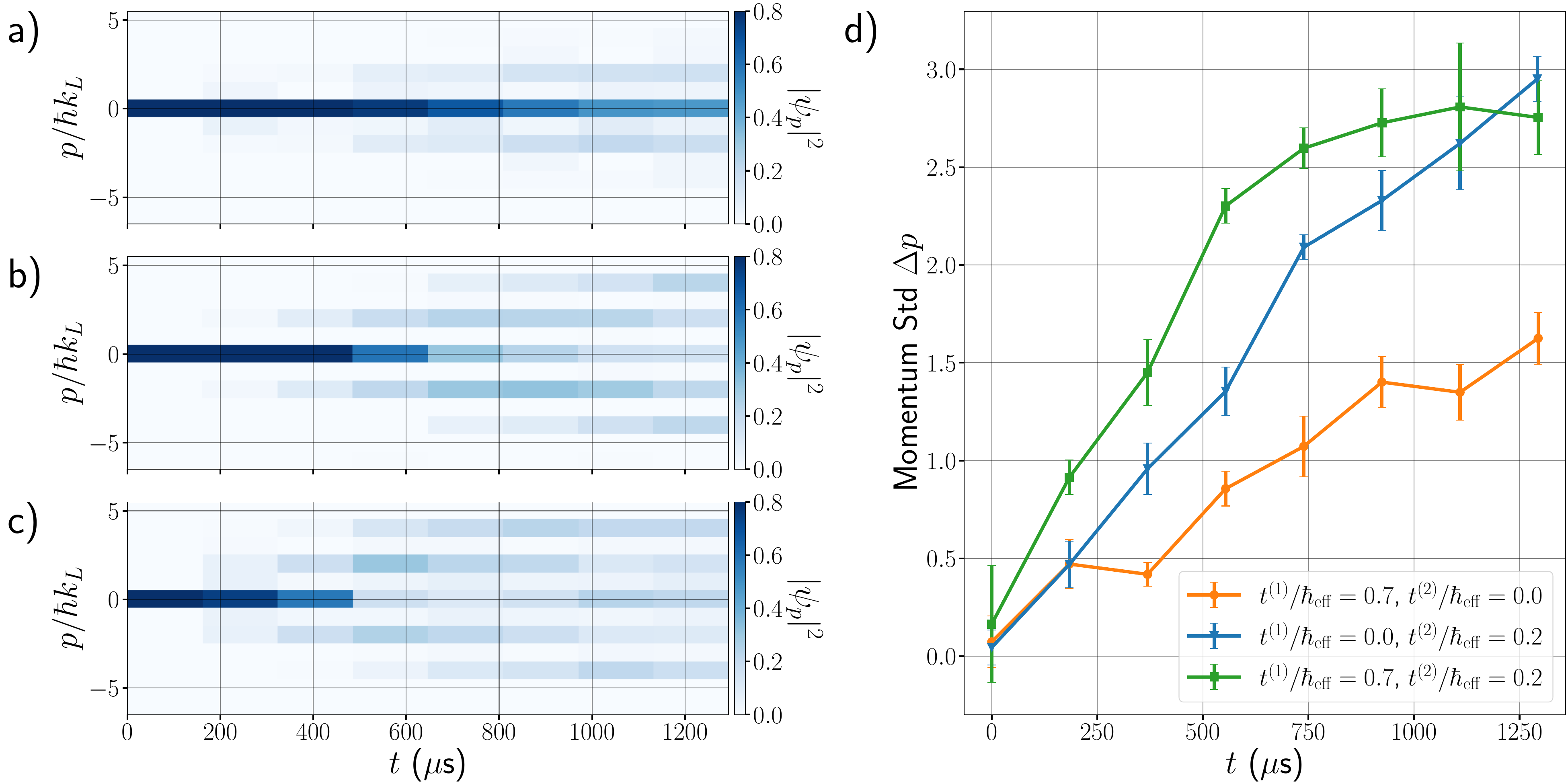}
	\caption{(a) Time evolution of the momentum distribution from the initial state $|0\rangle$ in a chain with  staggered onsite energies $\varepsilon_{2n} = 0$ and $\varepsilon_{2n+1}/\hbar_{\mathrm{eff}} = \pi$, and  nearest-neighbor tunneling $t^{(1)} = 0.7\,\hbar_{\mathrm{eff}}$, and no second-nearest-neighbor coupling ($t^{(2)} = 0$).  
(b) Time evolution of the momentum distribution in a chain with $t^{(1)} = 0$ and $t^{(2)} = 0.2\,\hbar_{\mathrm{eff}}$.  
(c) Time evolution of the momentum distribution in a chain with both $t^{(1)} = 0.7\,\hbar_{\mathrm{eff}}$ and $t^{(2)} =0.2\,\hbar_{\mathrm{eff}}$.  
(d) Time evolution of the wave-packet momentum width width $\Delta p = \sqrt{\langle \hat{p}^2\rangle - \langle \hat{p}\rangle^2}/(\hbar k_L)$. The Floquet period is $T_{6\pi} \approx 185.0\,\mu$s and the optimal control parameters used to obtain the modulation function are $M_T=1000$, $N=21$ and $N_t=5$.
}
	\label{Figure_Second_tunneling_coef_6pi_exp}
\end{figure*}

To investigate the role of the second-nearest-neighbor tunneling coefficient, we measure the evolution of an initially localized state $|0\rangle$ in three different tight-binding configurations at the $\hbar_{\rm eff}=6\pi$ resonance. The corresponding momentum distribution evolutions and the width of the wave function, defined as $\Delta p = \sqrt{\langle \hat{p}^2\rangle - \langle \hat{p}\rangle^2}/(\hbar k_L)$, are shown in Fig.~(\ref{Figure_Second_tunneling_coef_6pi_exp}).
In the first configuration (see Fig.~(\ref{Figure_Second_tunneling_coef_6pi_exp})a)), the nearest-neighbor tunneling amplitude is uniform along the chain, $t^{(1)} = 0.7 \times \hbar_{\mathrm{eff}}$, while the second-nearest-neighbor coupling is absent, $t^{(2)} = 0$. The onsite energies are staggered with period two: $\varepsilon_{2n} = 0$ and $\varepsilon_{2n+1} /\hbar_{\mathrm{eff}}= \pi $. 
The resulting dynamics is ballistic, as evidenced by the linear increase of $\Delta p$ with time. The wave packet spreads unequally over both even and odd sites, which originates from the energy offset between neighboring sites, which partially suppresses tunneling.
In the second configuration, see Fig.~(\ref{Figure_Second_tunneling_coef_6pi_exp})b), the nearest-neighbor coupling is turned off, $t^{(1)} = 0$, while a uniform second-nearest-neighbor tunneling is introduced, $t^{(2)} = 0.2 \times \hbar_{\mathrm{eff}}$. The onsite energies remain unchanged. 
Because second-nearest-neighbor tunneling connects only sites of identical parity, the dynamics is restricted to either the even or the odd sublattice. Since all sites within a given parity class have the same onsite energy, there is no energetic suppression of tunneling. 
As a result, the ballistic expansion speed is larger than in the first configuration, despite the smaller value of the tunneling amplitude. 
Finally, in the third configuration, see Fig.~(\ref{Figure_Second_tunneling_coef_6pi_exp})c), both tunneling processes are present, $t^{(1)} = 0.7 \times \hbar_{\mathrm{eff}}$ and $t^{(2)} = 0.2 \times \hbar_{\mathrm{eff}}$. 
In this case, the wave packet spreads even more rapidly than in the two previous situations, through the combination of couplings. The initial expansion is initially linear in time, before saturating when the wave packet reaches the boundaries of the finite chain.

\section{Discussion}

We have shown that the quantum resonances of a shaken rotor provide a generalized route to momentum-space lattice engineering, beyond the standard Bragg transitions approach.
However, the method presented in this work is subject to several limitations, which we now examine. These can be divided into two categories: limitations stemming from our experimental setup, which may be mitigated through technical improvements, and fundamental limitations inherent to the method itself.

\subsection{Experimental limits}\label{Limits_exp}

A first limitation arises from the finite quasi-momentum width of the atomic cloud. 
The resonant condition underlying our method requires $\beta_p = 0$. 
However, the Bose–Einstein condensate has a finite spatial extent and therefore a non-zero spread in quasi-momentum $\Delta \beta_p$. Although this width is difficult to measure directly, comparison between numerical simulations and experimental data indicates that it lies in the approximate range $\Delta \beta_p \in [0.01,\,0.05]\,\hbar k_L$.
One can estimate the timescale over which the approximation $\Delta \beta_p \approx 0$ remains valid using the following argument. 
Consider a MSL tight-binding model comprising $2N+1$ sites. 
Neglecting tunneling and modulation effects, we focus only on the kinetic energy contribution. A state located at the edge of the chain, with a quasi-momentum offset $\delta \beta_p$, will accumulate over a time $T$ a relative phase shift with respect to the component of the same order at $\beta_p=0$:
$\Delta \phi 
= \frac{(\hbar k_L N + \delta \beta_p)^2 T}{2m\hbar}
  - \frac{(\hbar k_L N)^2 T}{2m\hbar}
\simeq \frac{\delta \beta_q N k_L}{m}\, T$.
The resonant description remains valid as long as this relative phase 
remains small, \emph{i.e.}, $\Delta \phi \ll 2\pi$, which yields the condition
$T \ll \frac{2\pi m}{\delta \beta_q N k_L}$. This dephasing is maximal for the extremal orders.
For a chain of size $N \sim 5$ and using the upper estimate 
$\delta \beta_q \sim 0.05\,\hbar k_L$, this corresponds to a characteristic 
timescale $T^* \approx 250\,\mu\mathrm{s}$. 
Beyond this time, the phase dispersion across the cloud leads to 
inhomogeneous dephasing. Experimentally, this manifests as a blurring of 
the expected coherent dynamics, as observed for instance at long times 
in the evolution of the Gaussian state during Bloch oscillations 
(see Fig.~(\ref{Figure_Bloch_Oscillation_Gaussian})b). Fortunately, this estimation of the coherence timescale is a lower bound. In our experiments, the dephasing due to the quasi-momentum width becomes an issue only after $\sim 1$ms. 

A second limitation arises from the finite frequency bandwidth of the 
phase and amplitude modulation functions. 
The number of required Fourier harmonics increases with the size of the 
simulated chain: larger lattices require higher frequency components. 
Experimentally, the phase and amplitude are controlled using arbitrary 
waveform generators driving acousto-optic modulators (AOMs). 
The effective bandwidth of this setup is approximately $1.5$~MHz. 
This constraint limits the faithfully implementable tight-binding models 
to MSLs of roughly 30 sites, corresponding to momentum states 
$|n\rangle$ with $n \in [-15, +15]$. Finally, throughout this work we have treated the dynamics as a 
single-particle problem, neglecting atom–atom interactions. 
Our gas is sufficiently dilute that interactions are weak, and we estimate 
the atomic density to be $\rho \sim 10^{13}$ cm$^{-3}$. 
Under these conditions, the interaction energy  remains small compared to 
the characteristic tunneling and modulation energy scales, 
justifying the single-particle approximation on the timescales considered.

\subsection{Limits of the method}

Beyond the experimental constraints discussed above, the method itself 
also presents intrinsic limitations that restrict the class of tight-binding 
models that can be faithfully simulated.
First, both the perturbative analysis underlying the effective Hamiltonian 
and the fidelity of the unitary evolution obtained through optimal control 
reveal constraints on the accessible energy scales. 
In dimensionless units, the tunneling amplitudes must remain small compared 
to $\hbar_{\mathrm{eff}}^2$. 
In practice, this means that for most quantum resonances the tunneling 
coefficients must be small compared to the characteristic onsite energy 
differences set by the resonance condition. 
As a consequence, only a restricted region of parameter space can be explored. 
For instance, in the Rice–Mele model, this limitation prevents us from 
accessing the full topological phase diagram and, in particular, from 
continuously tuning parameters across the regime required to directly probe 
the Zak phase transition and the physics of Thouless pumps \cite{thouless_quantization_1983, nakajima_topological_2016, lohse_thouless_2016, citro_thouless_2023}.

Second, the onsite energies are not freely tunable parameters: they are fixed 
by the choice of quantum resonance. 
This lack of independent control prevents the realization of arbitrary 
onsite-energy landscapes. 
In particular, it is not possible to engineer topological defects through 
local modifications of the onsite energy, nor to implement disordered models 
such as the Anderson model with fully random onsite potentials.
Finally, if interactions between atoms are taken into account, the mapping 
onto standard lattice models becomes less straightforward. 
While interactions are local in real space, they are intrinsically nonlocal 
in momentum space \cite{An_PRL_2018}. 
As a result, the system does not naturally realize a simple Hubbard-type model, 
where interactions are onsite in the lattice basis. 
However, this feature could instead be viewed as an opportunity: 
momentum-space lattices provide a natural platform for simulating 
tight-binding models with long-range (all-to-all) interactions.

\subsection{Perspectives}\label{Perspectives}

A first natural perspective of this work would be to extend the method of quantum resonances combined with optimal control to higher dimensions. It is well know that most interesting condensed matter phenomena and models requires at least two space dimensions. 
Although the 2D kicked rotor resonances have not yet been studied, the Bragg resonance formalism has already been successfully extended to 2D, with the experimental study of chiral currents \cite{An_ScienceAdv_2017}. 
In that context, our work finds promising perspectives in recent proposals to simulate a wide variety of lattices \cite{Agrawal_MSL2D}, or to realize and detect topological phases in 2D MSLs  \cite{Wang_MSL_2D}.

Additionally the method developed in this work relies on two hypotheses: the unperturbed energy spectrum of the system is quadratic and it is experimentally possible to control the couplings between nearest quantum states. We can thus expect to extend this method to any system presenting this feature, like internal atomic sublevels with a quadratic Stark effect, coupled by laser-driven Raman transitions.

Finally, we restricted our optimal control approach to the quantum resonances of the kicked rotor because it is well-suited to the simulation of a tight-binding model with non zero on site energies. However, one can envision to use optimal control away from quantum resonances to fully design the unitary evolution operator.  Working in small Hilbert subspaces (from $\sim 2$ to $6$ momentum states), we recently demonstrated the realization of quantum gates with this method \cite{Flament_GATES_BEC}.

\paragraph*{Acknowledgements.} This work was supported by the Institut Universitaire de France, the ANR project QuCoBEC (ANR-22-CE47-0008) and the ANR PEPR Quantique QUTISYM (ANR-23-PETQ-0002). 

\appendix

\section{Kernel derivation}\label{annexe1}
Starting from the expression for the kernel from Eq.~\eqref{master_eq}:

\begin{align*}
	K(z,t) =& \sum_{k\in\mathbb{Z}}\frac{1}{2} e^{-i2\pi \frac{p}{q}(2k+1)t+i2\pi kz}\\
	=& \frac{1}{2}e^{-i2\pi\frac{p}{q}t}\sum_{k\in\mathbb{Z}}e^{i2\pi k(z-2\frac{p}{q}t)}\\
\end{align*}

We make use of the identity:
\begin{align*}
\sum_{k\in\mathbb{Z}}e^{i2\pi k z}=\sum_{\ell\in\mathbb{Z}}\delta (z-\ell)
\end{align*}
to get:
\begin{align*}
	K(z,t) =& \frac{1}{2}e^{-i2\pi\frac{p}{q}t}\sum_{\ell\in\mathbb{Z}}\delta(z-2\frac{p}{q}t-\ell)\\
	=& \frac{q}{4p}e^{-i2\pi\frac{p}{q}t}\sum_{\ell\in\mathbb{Z}}\delta\left(t-\frac{q}{2p}(z-\ell)\right)\\
	=&\frac{q}{4p}\sum_{\ell\in\mathbb{Z}}e^{-i\pi(z-\ell)}\delta\left(t-\frac{q}{2p}(z-\ell)\right)
\end{align*}
and we finally get equation \eqref{kernel}.

\section{Solution for $\hbar_{\mathrm{eff}} = 3\pi$}\label{annexe2}

We detail here the resolution for the modulation function in the quantum resonance $\hbar_{\mathrm{eff}} = 3\pi$. The method outlined here also applies to the resonance $\hbar_{\mathrm{eff}} = 6\pi$, recovering the result given in the main text.

Having first rewritten the equation on the modulation as Eq~(\ref{eq_fond_3pi}), we split the interval $[0,1]$ into $3$ parts, $[0,1/3]$, $[1/3,2/3]$ and $[2/3,1]$ and for $z\in[0,1/3]$ we define three functions $$f_{0}(z) = f(z),\, f_{1}(z) = f\left(z+\frac{2}{3}\right), \, f_{2}(z) = f\left(z+\frac{1}{3}\right).$$ Because of the periodicity of $f$, Eq~(\ref{eq_fond_3pi}) can be recast into a linear system, for $ z\in[0,1/3]$,

\begin{equation*}
\begin{cases}
     \tilde{T}(z) =& \alpha_0(z)f^0(z) +\alpha_1(z)f^1(z) +\alpha_2(z)f^2(z) \\
     \tilde{T}\left(z+\frac{1}{3}\right) =& \alpha_0\left(z+\frac{1}{3}\right)f^2(z) +\alpha_1\left(z+\frac{1}{3}\right)f^0(z)\\ &+\alpha_2\left(z+\frac{1}{3}\right)f^1(z) \\
     \tilde{T}\left(z+\frac{2}{3}\right) =& \alpha_0\left(z+\frac{2}{3}\right)f^1(z) +\alpha_1\left(z+\frac{2}{3}\right)f^2(z)\\ &+\alpha_2\left(z+\frac{2}{3}\right)f^0(z)
\end{cases}
\end{equation*}

We can now express the functions $\alpha_j(z)$ and the system then simplifies to
\begin{equation}
\begin{cases}
     \tilde{T}(z) &= f^0(z) -f^1(z)\\
     \tilde{T}\left(z+\frac{1}{3}\right) &= f^2(z) \\
     \tilde{T}\left(z+\frac{2}{3}\right) &= f^1(z) -f^0(z)
\end{cases}
\end{equation}
The first and last equation are in fact equivalent because $\tilde{T}\left(z+\frac{2}{3}\right) = -\tilde{T}(z)$. As a consequence, the systems does not have a unique solution and we can write down the general solution as
\begin{equation}
\begin{cases}
     f^0(z) &= a\tilde{T}(z)\\
     f^1(z) &= (a-1)\tilde{T}(z)\\
     f^2(z) &= \tilde{T}(z+\frac{1}{3})
\end{cases}
\end{equation}
with $a$ a complex number. The final expression of $f$ over the interval $[0,1]$ is given by the concatenation of the three solutions $f^0,\, f^1,\, \text{and}\, f^2$. In the symmetric case, when $a=1/2$ this leads to the expression Eq~(\ref{f_3pi}).
We can finally check that this function is indeed solution of the integral Eq.~(\ref{master_eq}).

\begin{align*}
    \int_0^1& e^{-i2\pi \frac{p}{q}(2k+1)t}\,f(t)\,dt\\
    &= \int_0^1 e^{-i \frac{3\pi}{2}(2k+1)t}\beta(t)e^{i \frac{3\pi}{2}t}\sum_n \Gamma_n t_n e^{3i\pi t} dt\\
    &=\sum_n \Gamma_n t_n \int_0^1\beta(t)e^{-i 3\pi(k-n)t}dt
\end{align*}

The function $\beta$ being piecewise constant, simple algebra provides the value of the integral:
\begin{align*}
    \int_0^1&\beta(t)e^{-i 3\pi(k-n)t}dt\\
    & = \int_0^{1/3}\frac{3}{2}e^{-i 3\pi(k-n)t}dt + \int_{1/3}^{2/3}3e^{-i 3\pi(k-n)t}dt \\
    & +\int_{2/3}^{1}\frac{3}{2}e^{-i 3\pi(k-n)t}dt\\
    & =2\delta_{k,n}
\end{align*}
which then yields as expected:
\begin{align*}
    \int_0^1& e^{-i2\pi \frac{p}{q}(2k+1)t}\,f(t)\,dt = 2\Gamma_k t_k.
\end{align*}

\medskip

\section{Gradient ascent method}\label{annexe3}

As the method consists in a gradient ascent method, we compute the gradient of the multi-time fidelity with respect to the modulation function (control parameter) $f_j^k$:
	
	\begin{align*}
		\frac{\partial \mathcal{F}}{\partial f_j^k} &= \frac{1}{N_t}\sum_{n=1}^{N_t}\frac{1}{N^2}\mathrm{tr}\left(\hat{U}_c^{\dagger^n}\hat{U}^{ ^n}(f_1,f_2)\right)^*\\
		&\times \mathrm{tr}\left(\hat{U}_c^{\dagger^n}\frac{\partial \hat{U}^{ ^n}(f_1,f_2)}{\partial f_j^k}\right) + c.c\\
		&= \frac{2}{N_t}\sum_{n=1}^{N_t}\frac{1}{N^2}\Re \left\{\mathrm{tr}\left(\hat{U}_c^{\dagger^n}\hat{U}^{ ^n}(f_1,f_2)\right)^*\right.\\
		&\left.\times \mathrm{tr}\left(\hat{U}_c^{\dagger^n}\frac{\partial \hat{U}^{ ^n}(f_1,f_2)}{\partial f_j^k}\right)\right\}
	\end{align*}
	
	\noindent We now have to take the derivative of the matrices product and we notice that each $f_j^k$ appears $n$ times:
	
	\begin{align*}
		\frac{\partial \hat{U}^{ ^n}(f_1,f_2)}{\partial f_j^k} &= 	\frac{\partial}{\partial f_j^k}\left(\prod_{k=1}^{M_T}e^{-\frac{i}{M_T\hbar_{\mathrm{eff}}}\hat{H}(f_1^k, f_2^k)}\right)^n\\
		&= -\frac{i}{M_T\hbar_{\mathrm{eff}}}\sum_{l=1}^n \hat{U}^{ ^{l-1}}(f_1,f_2) \prod_{m=k+1}^{M_T}e^{-\frac{i}{M_T\hbar_{\mathrm{eff}}}\hat{H}_m}\\
		&\times\frac{\partial \hat{H}(f_1^k,f_2^k)}{\partial f_j^k} \prod_{m=1}^{k-1}e^{-\frac{i}{M_T\hbar_{\mathrm{eff}}}\hat{H}_m} \hat{U}^{ ^{n-l}}(f_1,f_2)
	\end{align*}
	
	\noindent with $\hat{H}_m = \hat{H}(f_1^m, f_2^m)$. To simplify, we introduce the forward evolution operator:
	$\hat{U}^F_j = \prod_{m=1}^{j}e^{-\frac{i}{M_T\hbar_{\mathrm{eff}}}\hat{H}_m}$ and the backward evolution operator: $\hat{U}^B_j = \prod_{m=j}^{M_T}e^{-\frac{i}{M_T\hbar_{\mathrm{eff}}}\hat{H}_m}$ (the product is still taken with increasing $m$ from right to left). The total gradient can thus be written:
	\begin{widetext}
	\begin{equation}
	    \frac{\partial \mathcal{F}}{\partial f_j^k} = \frac{2}{M_TN_t N^2\hbar_{\mathrm{eff}}}\sum_{n=1}^{N_t}\sum_{l=1}^n \Im  \left\{\mathrm{tr}\left(\hat{U}_c^{\dagger^n}\hat{U}^{ ^n}(f_1,f_2)\right)^*\mathrm{tr}\left(\hat{U}_c^{\dagger^n}\hat{U}^{ ^{l-1}}(f)\hat{U}^B_{k+1}\frac{\partial \hat{H}(f_1^k,f_2^k)}{\partial f_j^k}\hat{U}^F_{k-1}\hat{U}^{ ^{n-l}}(f) \right)\right\}
	\end{equation}
\end{widetext}

\bibliography{references.bib}

\end{document}